\documentclass[11pt, oneside]{article}

\usepackage[english]{babel}
\usepackage{geometry}  
\usepackage{graphicx}      
\usepackage{tabularx}      
\usepackage{amssymb}
\usepackage{amsmath}

\usepackage{xltabular}
\usepackage[hyphens]{url}
\usepackage[hidelinks]{hyperref}
\hypersetup{breaklinks=true}
\usepackage[backend=biber,style=nature, doi=false,isbn=false, date=year, url=false, autopunct=true]{biblatex}
\makeatletter
\renewbibmacro*{textcite}{%
  \iffieldequals{namehash}{\cbx@lasthash}
    {\usebibmacro{cite:comp}}
    {\usebibmacro{cite:dump}%
     \ifbool{cbx:parens}
       {\bibclosebracket\global\boolfalse{cbx:parens}}
       {}%
     \iffirstcitekey
       {}
       {\textcitedelim}%
     \usebibmacro{cite:init}%
     \ifnameundef{labelname}
       {\printfield[citetitle]{labeltitle}}
       {\printnames{labelname}}%
     \addspace
     \ifnumequal{\value{citecount}}{1}
       {\usebibmacro{prenote}}
       {}%
     \mkbibsuperscript{\usebibmacro{cite:comp}}%
     \stepcounter{textcitecount}%
     \savefield{namehash}{\cbx@lasthash}}}
\makeatother

\usepackage{csquotes}
\usepackage{cleveref}
\usepackage{siunitx}
\usepackage{geometry}
\usepackage{multirow}
\usepackage[dvipsnames,table,xcdraw]{xcolor}
\definecolor{tumblue}{rgb}{0.0980392157,0.2901960784,0.5058823529}

\usepackage{xspace}
\usepackage{caption}
\usepackage{subcaption}
\usepackage{authblk}

\usepackage{marvosym}

\usepackage{sectsty} 

\sectionfont{\color{tumblue}\selectfont\sffamily}
\subsectionfont{\color{tumblue}\selectfont\sffamily}
\subsubsectionfont{\color{tumblue}\selectfont\sffamily}
\paragraphfont{\selectfont\sffamily}
\subparagraphfont{\color{gray}\selectfont\sffamily}

\definecolor{halfgray}{gray}{0.55}

\IfFileExists{./fonts/Heliotrope-Bold.otf}{
    \usepackage{fontspec}
    \setmainfont{Heliotrope}[Extension = .otf, UprightFont = *-Regular, ItalicFont = *-Italic, BoldFont=*-Bold, Path = ./fonts/]
}{
    \usepackage{biolinum}
    
}

\usepackage{lipsum}
\usepackage[labelfont=bf]{caption}
\usepackage[nonumberlist,acronym]{glossaries}
\newglossaryentry{cotg}{name={Chain-of-Thought}, description={A prompt engineering technique that involves a series of intermediate natural language reasoning steps that lead to the final output.\autocite{wei2023chainofthought} CoT encourages an \gls{llm} to explain its reasoning step by step (e.g., by prompting it to \enquote{think step by step}), and it is intended to improve the ability of \glspl{llm} to perform complex reasoning}}

\newglossaryentry{coveg}{name={Chain-of-Verification}, description={A four-step process intended to reduce hallucination in \glspl{llm}:\autocite{dhuliawala2023chainofverification} (i) the \gls{llm} generates a baseline response; (ii) generates a list of verification questions to self-analyze if there are any mistakes in the original response; (iii) answers each verification question in turn, checking the answer against the original response to check for inconsistencies or mistakes; and (iv) generates the final revised response incorporating the verification results}}

\newglossaryentry{ieg}{name={Information extraction}, description={The process of automatically extracting information from unstructured text. This process involves converting raw text into structured data by recognizing entities, relationships, and events}}

\newglossaryentry{llmg}{name={Large language model}, description={A language model that has a large number of parameters. There is no agreed rule for the number of parameters that makes a language model large enough to be called a large language model, but this number is usually on the scale of billions.\autocite{minaee2024} For example, Llama 3, GPT-3, and GPT-4 contain 70B, 175B, and 1.76T parameters, respectively, while Claude 3 Opus is estimated to have 2 trillion parameters. Most current \glspl{llm} are based on the Transformer architecture. \Glspl{llm} can perform many types of language tasks, such as generating human-like text, understanding context, translation, summarization, and question-answering}}

\newglossaryentry{lmg}{name={Language model}, description={A \gls{ml} model that estimates the probability of a token or sequence of tokens occurring in a longer sequence of tokens. This probability is used to predict the most likely next token based on the previous sequence. Language models are trained on large datasets of text, learning the patterns and structures of language to understand, interpret, and generate natural language}}

\newglossaryentry{lorag}{name={Low-rank adaptation}, description={A \gls{peft} technique that freezes the pre-trained model weights and decomposes the weights update matrix into two lower-rank matrices that contain a reduced number of trainable parameters to be optimized during fine-tuning\autocite{hu2021lora}}}

\newglossaryentry{mlmg}{name={Masked language modeling}, description={A method of self-supervised learning in which a random sample of the input tokens are masked and, during the training, the model is asked to predict the original inputs for the masked tokens based on the context provided by the unmasked tokens}}

\newglossaryentry{nerg}{name={Named entity recognition}, description={A \gls{nlp} technique that identifies and classifies named entities within text into predefined categories such as names of people, organizations, locations, dates, and other entities}}

\newglossaryentry{nerreg}{name={Name entity recognition and relation extraction}, description={A combined process in \gls{nlp} that first identifies and categorizes entities (such as materials and properties) within a text and then determines semantic relationships between these recognized entities. This dual approach enables the extraction of structured data, revealing how entities interact within the text, and is crucial for building knowledge graphs or information retrieval systems}}

\newglossaryentry{nlpg}{name={Natural language processing}, description={A subfield of computer science that uses machine learning to enable computers to understand, interpret, and generate human language. The main tasks in NLP are speech recognition, text classification, natural-language understanding, and natural-language generation}}

\newglossaryentry{nspg}{name={Next sentence prediction}, description={A task used during the pre-training of \glspl{llm} where the model is trained to predict whether a given sentence logically follows another sentence. This task helps the model learn the relationships between sentences, improving its understanding of context and coherence in text. By training on vast amounts of text data, the model develops the ability to generate coherent and contextually appropriate continuations for a given text}}

\newglossaryentry{ocrg}{name={Optical character recognition}, description={A technique used to identify and convert images of printed or handwritten text into a machine-readable text format. This involves segmentation of text regions, character recognition, and post-processing to correct errors and enhance accuracy}}

\newglossaryentry{peftg}{name={Parameter-efficient fine-tuning}, description={A methodology to efficiently fine-tune large pre-trained models without modifying their original parameters. PEFT strategies involve adjusting only a small number of additional model parameters during fine-tuning on a new, smaller training dataset. This significantly reduces the computational and storage costs while achieving comparable performance to full fine-tuning}}

\newglossaryentry{ppog}{name={Proximal policy optimization}, description={A reinforcement learning algorithm used to train \glspl{llm} for alignment by optimizing their behavior through interaction with an environment, balancing exploration and exploitation. PPO achieves this by adjusting the policy parameters in a way that keeps changes within a predefined safe range to maintain stability and improve learning efficiency. This method helps ensure that the \gls{llm} aligns its outputs with desired goals, such as ethical guidelines or user intentions, by iteratively refining its responses based on feedback. PPO is often used as part of \gls{rlhf}}}

\newglossaryentry{ragg}{name={Retrieval augmented generation}, description={A technique for improving the quality of text generation by providing \glspl{llm} with access to information retrieved from external knowledge sources. In practice, this means that relevant retrieved text snippets are added to the prompt}}

\newglossaryentry{reg}{name={Relation extraction}, description={A \gls{nlp} task that identifies and categorizes the semantic relationships between entities within a text, e.g., materials and properties}}

\newglossaryentry{rlhfg}{name={Reinforcement learning from human feedback}, description={A mechanism to align \glspl{llm} with user preferences by fine-tuning with human feedback. Users are asked to rate the quality of a model's response. Based on this, a preference model is trained and then used in a reinforcement learning setup to optimize the generations of the \gls{llm}}}

\newglossaryentry{ropeg}{name={Rotational positional encoding}, description={Rotational positional encoding is a method used in \glspl{llm} to represent the position of tokens within a sequence. Instead of using absolute positional values, this technique encodes positional information as rotations in a continuous vector space, which allows the model to understand both the absolute position of tokens and their relative distances.}}

\newglossaryentry{sftg}{name={Supervised fine-tuning}, description={The process in which a pre-trained \gls{llm} is fine-tuned on a labeled dataset of a specific task. It involves adapting the parameters of the pre-trained model to improve its performance on the new task, leveraging the knowledge and representations learned from the initial training}}

\newglossaryentry{sslg}{name={Self-supervised learning}, description={A machine learning technique that involves generating labels from the input data itself instead of relying on external labeled data. It has been foundational for the success of \glspl{llm}, as their pre-training task (next word prediction or filling in of masked words) is a self-supervised task}}

\newglossaryentry{vdug}{name={Visual document understanding}, description={Capability of systems to automatically interpret and analyze the content of visual documents, such as images, scanned papers, or digital documents containing complex layouts and graphics. It involves employing computer vision techniques to extract textual information, recognize structural elements like tables and diagrams, and comprehend the overall content}}

\newglossaryentry{vlmg}{name={Vision language model}, description={A multimodal model that can learn simultaneously from images and texts, generating text outputs}}
\makenoidxglossaries 
\setacronymstyle{long-short}
\newglossaryentry{ann}{
    type=\acronymtype, 
    name={ANN}, 
    description={artificial neural network},
    first={artificial neural network (ANN)}, 
    firstplural={artificial neural networks (ANNs)}
}

\newglossaryentry{api}{
    type=\acronymtype, 
    name={API}, 
    description={Application Programming Interface},
    first={Application Programming Interface (API)}, 
    firstplural={Application Programming Interfaces (APIs)}
}

\newglossaryentry{bfs}{
    type=\acronymtype, 
    name={BFS}, 
    description={breadth-first search},
    first={breadth-first search (BFS)}, 
    firstplural={breadth-first searches (BFSs)}
}

\newglossaryentry{cli}{
    type=\acronymtype, 
    name={CLI}, 
    description={Command-Line Interface},
    first={Command-Line Interface (CLI)}, 
    firstplural={Command-Line Interfaces (CLIs)}
}

\newglossaryentry{cot}{
    type=\acronymtype, 
    name={CoT}, 
    description={Chain-of-Thought}, 
    see=[Glossary:]{cotg},
    first={Chain-of-Thought (CoT)}, 
    firstplural={Chain-of-Thoughts (CoTs)}
}

\newglossaryentry{cove}{
    type=\acronymtype, 
    name={CoVe}, 
    description={Chain-of-Verification}, 
    see=[Glossary:]{coveg},
    first={Chain-of-Verification (CoVe)}, 
    firstplural={Chain-of-Verifications (CoVes)}
}

\newglossaryentry{dfs}{
    type=\acronymtype, 
    name={DFS}, 
    description={depth-first search},
    first={depth-first search (DFS)}, 
    firstplural={depth-first searches (DFSs)}
}

\newglossaryentry{eae}{
    type=\acronymtype, 
    name={EAE}, 
    description={event argument extraction},
    first={event argument extraction (EAE)}, 
    firstplural={event argument extractions (EAEs)}
}

\newglossaryentry{FN}{
    type=\acronymtype, 
    name={FN}, 
    description={false negative},
    first={false negative (FN)}, 
    firstplural={false negatives (FNs)}
}

\newglossaryentry{FP}{
    type=\acronymtype, 
    name={FP}, 
    description={false positive},
    first={false positive (FP)}, 
    firstplural={false positives (FPs)}
}

\newglossaryentry{html}{
    type=\acronymtype, 
    name={HTML}, 
    description={hypertext markup language},
    first={hypertext markup language (HTML)}, 
    firstplural={hypertext markup languages (HTMLs)}
}

\newglossaryentry{ie}{
    type=\acronymtype, 
    name={IE}, 
    description={information extraction}, 
    see=[Glossary:]{ieg},
    first={information extraction (IE)}, 
    firstplural={information extractions (IEs)}
}

\newglossaryentry{llm}{
    type=\acronymtype, 
    name={LLM}, 
    description={large language model}, 
    see=[Glossary:]{llmg},
    first={large language model (LLM)}, 
    firstplural={large language models (LLMs)}
}

\newglossaryentry{lm}{
    type=\acronymtype, 
    name={LM}, 
    description={language model}, 
    see=[Glossary:]{lmg},
    first={language model (LM)}, 
    firstplural={language models (LMs)}
}

\newglossaryentry{lora}{
    type=\acronymtype, 
    name={LoRA}, 
    description={low-rank adaptation}, 
    see=[Glossary:]{lorag},
    first={low-rank adaptation (LoRA)}, 
    firstplural={low-rank adaptations (LoRAs)}
}

\newglossaryentry{mae}{
    type=\acronymtype, 
    name={MAE}, 
    description={mean absolute error},
    first={mean absolute error (MAE)}, 
    firstplural={mean absolute errors (MAEs)}
}

\newglossaryentry{ml}{
    type=\acronymtype, 
    name={ML}, 
    description={machine learning},
    first={machine learning (ML)}, 
    firstplural={machine learnings (MLs)}
}

\newglossaryentry{mlm}{
    type=\acronymtype, 
    name={MLM}, 
    description={masked language modeling}, 
    see=[Glossary:]{mlmg},
    first={masked language modeling (MLM)}, 
    firstplural={masked language modelings (MLMs)}
}

\newglossaryentry{mof}{
    type=\acronymtype, 
    name={MOF}, 
    description={metal-organic framework},
    first={metal-organic framework (MOF)}, 
    firstplural={metal-organic frameworks (MOFs)}
}

\newglossaryentry{ner}{
    type=\acronymtype, 
    name={NER}, 
    description={named entity recognition}, 
    see=[Glossary:]{nerg},
    first={named entity recognition (NER)}, 
    firstplural={named entity recognitions (NERs)}
}

\newglossaryentry{nerre}{
    type=\acronymtype, 
    name={NERRE}, 
    description={name entity recognition and relation extraction}, 
    see=[Glossary:]{nerreg},
    first={name entity recognition and relation extraction (NERRE)}, 
    firstplural={name entity recognitions and relation extractions (NERREs)}
}

\newglossaryentry{nlp}{
    type=\acronymtype, 
    name={NLP}, 
    description={natural language processing}, 
    see=[Glossary:]{nlpg},
    first={natural language processing (NLP)}, 
    firstplural={natural language processings (NLPs)}
}

\newglossaryentry{nmr}{
    type=\acronymtype, 
    name={NMR}, 
    description={nuclear magnetic resonance},
    first={nuclear magnetic resonance (NMR)}, 
    firstplural={nuclear magnetic resonances (NMRs)}
}

\newglossaryentry{nn}{
    type=\acronymtype, 
    name={NN}, 
    description={neural network},
    first={neural network (NN)}, 
    firstplural={neural networks (NNs)}
}

\newglossaryentry{nsp}{
    type=\acronymtype, 
    name={NSP}, 
    description={next sentence prediction}, 
    see=[Glossary:]{nspg},
    first={next sentence prediction (NSP)}, 
    firstplural={next sentence predictions (NSPs)}
}

\newglossaryentry{ocr}{
    type=\acronymtype, 
    name={OCR}, 
    description={optical character recognition}, 
    see=[Glossary:]{ocrg},
    first={optical character recognition (OCR)}, 
    firstplural={optical character recognitions (OCRs)}
}

\newglossaryentry{peft}{
    type=\acronymtype, 
    name={PEFT}, 
    description={parameter-efficient fine-tuning}, 
    see=[Glossary:]{peftg},
    first={parameter-efficient fine-tuning (PEFT)}, 
    firstplural={parameter-efficient fine-tunings (PEFTs)}
}

\newglossaryentry{ppo}{
    type=\acronymtype, 
    name={PPO}, 
    description={proximal policy optimization}, 
    see=[Glossary:]{ppog},
    first={proximal policy optimization (PPO)}, 
    firstplural={proximal policy optimizations (PPOs)}
}

\newglossaryentry{rag}{
    type=\acronymtype, 
    name={RAG}, 
    description={retrieval augmented generation}, 
    see=[Glossary:]{ragg},
    first={retrieval augmented generation (RAG)}, 
    firstplural={retrieval augmented generations (RAGs)}
}

\newglossaryentry{re}{
    type=\acronymtype, 
    name={RE}, 
    description={relation extraction}, 
    see=[Glossary:]{reg},
    first={relation extraction (RE)}, 
    firstplural={relation extractions (REs)}
}

\newglossaryentry{rlhf}{
    type=\acronymtype, 
    name={RLHF}, 
    description={reinforcement learning from human feedback}, 
    see=[Glossary:]{rlhfg},
    first={reinforcement learning from human feedback (RLHF)}, 
    firstplural={reinforcement learnings from human feedback (RLHFs)}
}

\newglossaryentry{rnn}{
    type=\acronymtype, 
    name={RNN}, 
    description={recurrent neural network},
    first={recurrent neural network (RNN)}, 
    firstplural={recurrent neural networks (RNNs)}
}

\newglossaryentry{rope}{
    type=\acronymtype, 
    name={RoPE}, 
    description={rotational positional encoding}, 
    see=[Glossary:]{ropeg},
    first={rotational positional encoding (RoPE)}, 
    firstplural={rotational positional encodings (RoPEs)}
}

\newglossaryentry{sft}{
    type=\acronymtype, 
    name={SFT}, 
    description={supervised fine-tuning}, 
    see=[Glossary:]{sftg},
    first={supervised fine-tuning (SFT)}, 
    firstplural={supervised fine-tunings (SFTs)}
}

\newglossaryentry{ssl}{
    type=\acronymtype, 
    name={SSL}, 
    description={self-supervised learning}, 
    see=[Glossary:]{sslg},
    first={self-supervised learning (SSL)}, 
    firstplural={self-supervised learnings (SSLs)}
}

\newglossaryentry{tdm}{
    type=\acronymtype, 
    name={TDM}, 
    description={text and data mining},
    first={text and data mining (TDM)}, 
    firstplural={text and data minings (TDMs)}
}

\newglossaryentry{TN}{
    type=\acronymtype, 
    name={TN}, 
    description={true negative},
    first={true negative (TN)}, 
    firstplural={true negatives (TNs)}
}

\newglossaryentry{TP}{
    type=\acronymtype, 
    name={TP}, 
    description={true positive},
    first={true positive (TP)}, 
    firstplural={true positives (TPs)}
}

\newglossaryentry{vdu}{
    type=\acronymtype, 
    name={VDU}, 
    description={visual document understanding}, 
    see=[Glossary:]{vdug},
    first={visual document understanding (VDU)}, 
    firstplural={visual document understandings (VDUs)}
}

\newglossaryentry{vlm}{
    type=\acronymtype, 
    name={VLM}, 
    description={vision language model}, 
    see=[Glossary:]{vlmg},
    first={vision language model (VLM)}, 
    firstplural={vision language models (VLMs)}
}

\newglossaryentry{xml}{
    type=\acronymtype, 
    name={XML}, 
    description={extensible markup language},
    first={extensible markup language (XML)}, 
    firstplural={extensible markup languages (XMLs)}
}

\makenoidxglossaries

 \usepackage{microtype}
 \usepackage{fontawesome}

\definecolor{linkcolor}{RGB}{0, 102, 204} 

\newcommand{\OnlineMaterial}[2]{%
  \href{#2}{\textcolor{linkcolor}{\faGlobe\, Online Material |  #1}}%
}

\newcommand{\nougat}{Nougat\xspace}
\newcommand{\marker}{Marker\xspace}

\newcommand{\gptfourvision}{GPT-4 Vision\xspace}
\newcommand{\gptfour}{GPT-4\xspace}
\newcommand{\gptthreefive}{GPT-3.5\xspace}

\newcommand{\gptthree}{GPT-3\xspace}
\newcommand{\claudethree}{Claude 3\xspace}

\newcommand{\llamathree}{Llama 3\xspace}
\newcommand{\llamathreeseventy}{Llama 3-70B\xspace}
\newcommand{\llamatwo}{Llama 2\xspace}
\newcommand{\llamatwoseven}{Llama 2-7B\xspace}
\newcommand{\geminipro}{Gemini Pro\xspace}

\usepackage{credits}
\usepackage{orcidlink}
\usepackage{hyperref}
\usepackage{multirow}
\usepackage{threeparttable}
\usepackage{arydshln}
\usepackage{xcolor}
\usepackage{booktabs}
\usepackage{tcolorbox}

\title{\textsf{From Text to Insight: Large Language Models for Chemical Data Extraction}}

\author[1, $\star$]{Mara~Schilling-Wilhelmi~\orcidlink{0009-0007-4392-5918}}
\author[1,2, $\star$]{Martiño~Ríos-García~\orcidlink{0000-0003-1507-4048}}
\author[3]{Sherjeel~Shabih~\orcidlink{0009-0007-4392-5918}}
\author[2]{María~Victoria~Gil~\orcidlink{0000-0002-2258-3011}}
\author[4]{Santiago~Miret~\orcidlink{0000-0002-5121-3853}}
\author[3]{Christoph~T.~Koch~\orcidlink{0000-0002-3984-1523}}
\author[3]{José~A.~Márquez\orcidlink{0000-0002-8173-2566}}
\author[1,5,6, \Letter]{Kevin~Maik~Jablonka~\orcidlink{0000-0003-4894-4660}}

\affil[1]{Laboratory of Organic and Macromolecular Chemistry (IOMC), Friedrich Schiller University Jena, Humboldtstrasse 10, 07743 Jena, Germany}
\affil[2]{Institute of Carbon Science and Technology (INCAR), CSIC, Francisco Pintado Fe 26, 33011 Oviedo, Spain}
\affil[3]{Department of Physics and CSMB, Humboldt-Universität zu Berlin, Berlin, Germany}
\affil[4]{Intel Labs}
\affil[5]{Center for Energy and Environmental Chemistry Jena (CEEC Jena), Friedrich Schiller University Jena, Philosophenweg 7a, 07743 Jena, Germany}
\affil[6]{Helmholtz Institute for Polymers in Energy Applications Jena (HIPOLE Jena), Lessingstrasse 12-14, 07743 Jena, Germany}

\affil[$\star$]{These authors contributed equally.}
\affil[\Letter]{\texttt{mail@kjablonka.com}}

\begin{document}
\maketitle
 
\begin{abstract}
\noindent The vast majority of chemical knowledge exists in unstructured natural language, yet structured data is crucial for innovative and systematic materials design. Traditionally, the field has relied on manual curation and partial automation for data extraction for specific use cases. 
The advent of large language models (LLMs) represents a significant shift, potentially enabling non-experts to extract structured, actionable data from unstructured text efficiently.
While applying LLMs to chemical and materials science data extraction presents unique challenges, domain knowledge offers opportunities to guide and validate LLM outputs. 
This tutorial review provides a comprehensive overview of LLM-based structured data extraction in chemistry, synthesizing current knowledge and outlining future directions. We address the lack of standardized guidelines and present frameworks for leveraging the synergy between LLMs and chemical expertise.
This work serves as a foundational resource for researchers aiming to harness LLMs for data-driven chemical research.  The insights presented here could significantly enhance how researchers across chemical disciplines access and utilize scientific information, potentially accelerating the development of novel compounds and materials for critical societal needs.
\end{abstract}

\pagebreak

\begin{tcolorbox}[]
\textbf{Key learning points}

\begin{enumerate}
\item End-to-end workflow for LLM-based chemical data extraction: from data collection to structured output

\item Advanced techniques to enhance extraction: multimodal approaches, and agentic systems

\item Quality assurance in chemical data extraction: constrained decoding, domain-specific validation

\item Future frontiers: addressing cross-document analysis, diverse modalities, and emerging challenges in chemical LLMs
\end{enumerate}
\end{tcolorbox}

\pagebreak
\tableofcontents

\pagebreak

\pagebreak
\section{Introduction} 

Materials and molecular design have a long history of using empirical correlations or models to guide the design and synthesis of new compounds---or to inspire new theories. 
For example, structured data has been used via Ashby plots to select materials,\autocite{Ashby1999-tb} or via scaling relations to design catalysts.\autocite{abild2007scaling}
More recently, structured data has been used via machine learning models to predict the properties of compounds before they have been synthesized.\autocite{Butler_2018, jablonka2020big, Ramprasad_2017,Choudhary2022} 
In some cases, models can even directly recommend materials based on desired properties or guide the design of chemical experiments.\autocite{raccuglia2016machine, sanchez2018inverse}
A foundation for all these data uses is that the data must be available in a structured form, e.g., a table with clearly defined columns (often representing specific chemical or physical properties). 

\begin{figure}
    \centering
    \includegraphics[width=\textwidth]{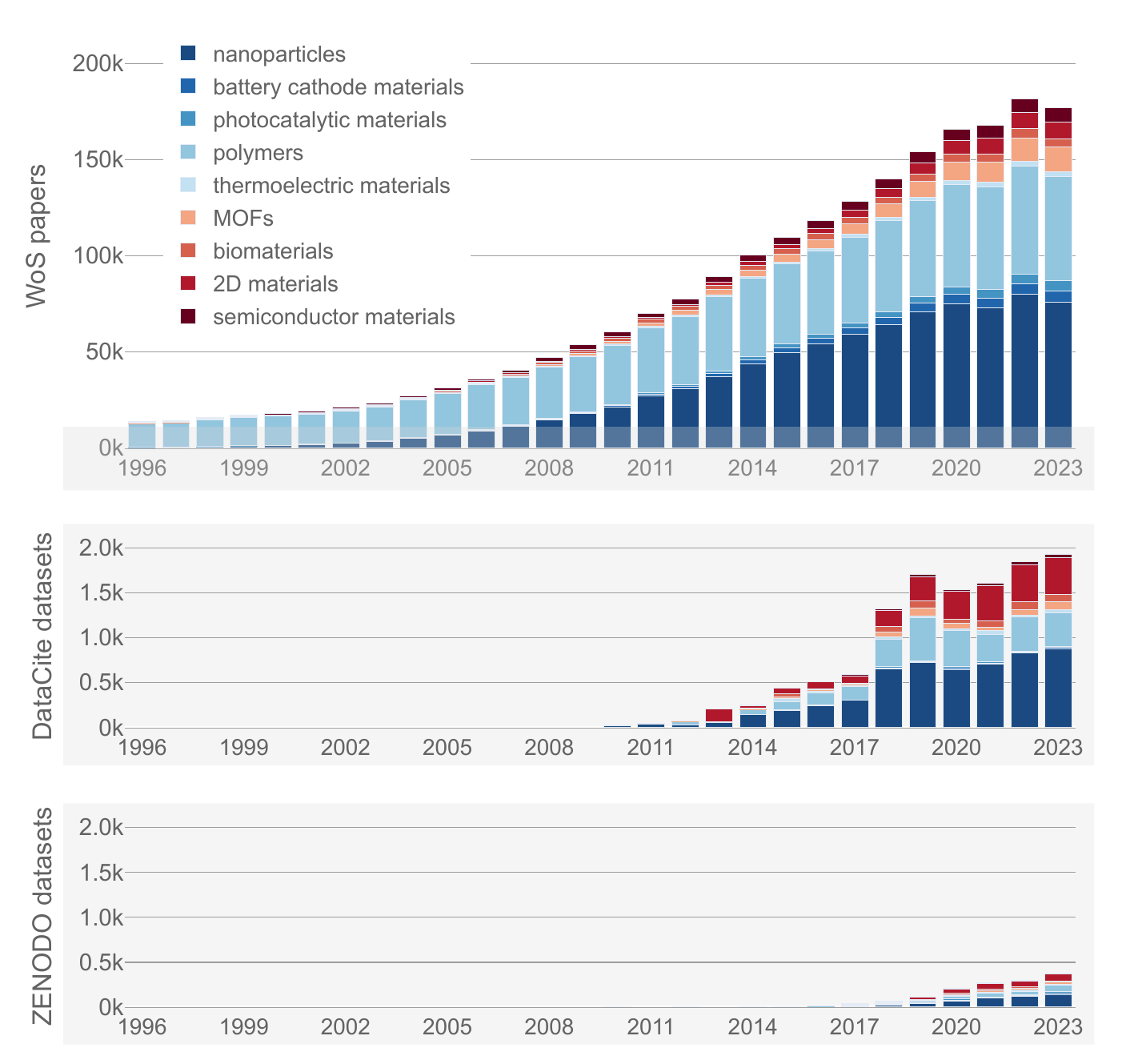}
    \caption{\textbf{Number of research papers vs. datasets deposited in data repositories in materials science and chemistry per year.} The top graph shows the number of publications from 1996 to 2023. The number of records was obtained from the search queries \enquote{(nanoparticles)}, \enquote{(battery AND cathode AND materials)}, \enquote{(photocatalytic AND materials)}, \enquote{(polymers)}, \enquote{(thermoelectric AND materials)}, \enquote{((metal-organic AND framework) OR MOF)}, \enquote{(biomaterials)}, \enquote{((2D AND materials) OR graphene)}, and \enquote{(semiconductor AND materials)} in the Web of Science Core Collection on July 1, 2024 (search based on title, abstract and indexing, including \enquote{Article} and \enquote{Data Paper} document types) (categories based on \textcite{Kononova_2021}) (see \OnlineMaterial{Research articles vs datasets in chemistry and materials science}{https://matextract.pub/content/intro_figure/figure1_intro_notebook.html}). The two graphs below show the number of datasets in chemistry and materials science deposited in the Zenodo and DataCite repositories from 1996 to 2023. The number of records was obtained from similar queries by restricting the document type to \enquote{Dataset}. Note the different $y$-axis scale between the top and bottom graphs. While this figure highlights the large difference in the availability of structured datasets compared to papers, we note that a one-to-one comparison of these numbers is not always fair. This is because sometimes multiple papers can be used to create a single dataset, as in the case of curated databases, or vice versa, where multiple structured datasets can result from a single paper's work.}
    \label{fig:struc_vs_paper}
\end{figure}

While there have been many advances in research data management, only a minuscule fraction of all available research data is available in a usable structured form (see \Cref{fig:struc_vs_paper}). 
Most data is still only reported in a compressed and highly processed form via text such as scientific articles. 
Thus, there is a large untapped potential in leveraging this unstructured text data.

The promise of leveraging data \enquote{hidden} in scientific articles---such as links between disjoint data published in separate articles (\enquote{Swanson links})---has motivated researchers to develop dedicated extraction pipelines.\autocite{Krallinger2017,Mavrai2021}  
These pipelines often relied on rule-based approaches,\autocite{ Jessop2011, Lowe_2015, Hawizy_2011, Swain2016, Mehr_2020} or smaller \gls{ml} models trained on manually annotated corpora.\autocite{Guo_2021, Rockt_schel_2012, Kononova_2019, Huang_2022, Shetty2023} 
Those approaches, however, face challenges with the diversity of topics and reporting formats in chemistry and materials research as they are hand-tuned for very specific use cases.\autocite{hira2024reconstructing} 
In 2019, researchers still struggled to effectively utilize data from old PDFs as   \enquote{currently available solutions still fail[ed] to provide high enough accuracy to reliably extract chemistry}\autocite{Kononova_2021} and the development of extraction workflows typically required very deep understanding and investment of multiple months of development time that had to be recommitted for every new use case. Data extraction, thus, presents a \enquote{death by 1000 cuts} problem. Automating the extraction for one particular case might not be too difficult, but the sheer scale of possible variations makes the overall problem intractable.\autocite{chrisre_keynote, borgman2016data} 

With advances in \glspl{llm} this has dramatically changed because \glspl{llm} can solve tasks for which they have not been explicitly trained. \autocite{vaswani2023attention,Yenduri2023, Song2019,Wu2024}
\glspl{llm} thus present a powerful and scalable alternative for structured data extraction. Using those \glspl{llm}, many data extraction workflows that would have taken weeks or longer to develop can now be bootstrapped into a prototype over the course of a two-day hackathon.\autocite{Jablonka_2023}

While there has been a growing number of reports showing the use of \glspl{llm} for data extraction in chemistry\autocite{Zhang2024chemicaltext,Smith2024enzimes,Li2024ChemVLM,chen2024coreferences} and materials science,\autocite{dagdelen_structured_2024,Shetty2023,Choi2024,lei_materials_2024,Polak_2024,ye_construction_2024,Polak2024b,Suvarna2023} there is still no clear framework on how to best build and evaluate such extraction pipelines for materials and molecular data. 
In many cases, chemical data extraction is more complicated than in other domains.\autocite{khalighinejad2024extracting} However, chemical expertise and physical laws also offer unique possibilities to constrain or validate the results.

This review aims to leverage the synergy between \glspl{llm} and chemistry and provide a practical guide for materials scientists and chemists looking to harness the power of \glspl{llm} for structured data extraction. We cover the entire workflow and real-world applications, drawing from the latest research and insights from the field. By bridging the gap between \gls{llm} research and practical application in chemistry and materials science, we aim to empower researchers to make the most of these powerful models.
To facilitate this, each section will be accompanied by a section in an executable Jupyter Book\autocite{ExecutableBooksCommunity_2020} that elaborates on hands-on examples (\OnlineMaterial{matextract.pub}{https://matextract.pub}).\autocite{matextract}

\section{Overview of the working principles of LLMs} \label{sec:background}

\begin{figure}[htbp]
    \centering
    \includegraphics[width=\textwidth]{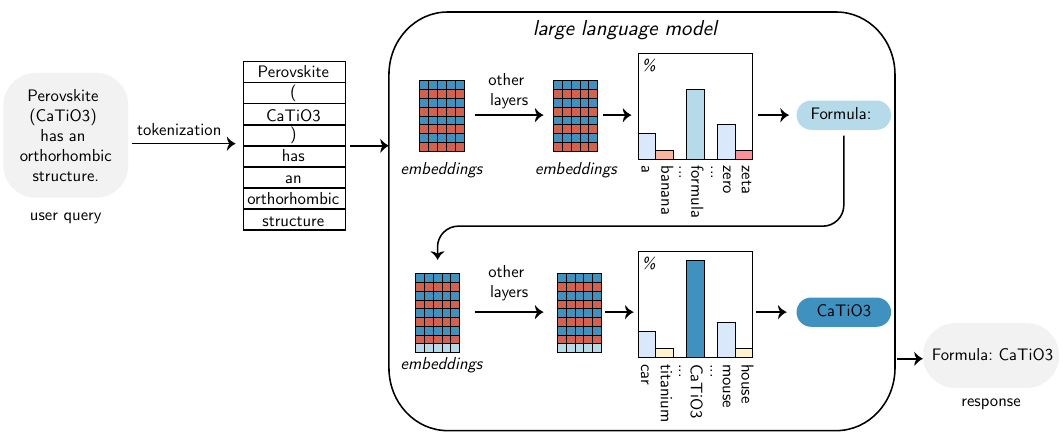}
    \caption{\textbf{High-level explanation of the working principle of an LLM}. The data flow in the image corresponds to a decoder-only model, e.g., a GPT or Llama model. One token is produced each time, considering all the tokens from the input and all previously produced tokens. The process starts with the tokenizer, which converts the user query into smaller constituent units, the tokens. The tokens are passed into the model, where input embeddings are computed, after which additional operations transform the embeddings. As a result, the model outputs the probabilities over possible subsequent tokens. Depending on the temperature parameter, the most probable token or a less probable one is chosen, and the sampled token is added to the input. By repeating this process, the model generates the response to the query.}
    \label{fig:background_image}
\end{figure}

\noindent 
At the most basic level, \glspl{llm} work by completing the text they receive as input with one of the (what they compute to be) most likely tokens (\Cref{fig:background_image}). 
In this context, tokens are the basic units of text on which those models operate. They are usually words but can be suffixes, prefixes, or individual characters. A model knows only a finite number of tokens, which is typically known as the vocabulary. This vocabulary must be determined prior to the training of the model and can limit the performance of models, e.g., via suboptimal splitting of chemical formulas or numbers.\autocite{singh2024tokenization} 

\paragraph{Sampling outputs} The output of the most commonly used models (so-called decoder-only models) is a probability distribution over tokens, i.e., a measure of how likely each token the model knows is as a completion of the input sequence (\Cref{fig:background_image}).
While it might seem most intuitive to choose the most likely token (so-called greedy decoding), this often leads to unnatural-sounding text. Thus, in some applications, it is common to choose less likely tokens. The frequency of this happening is determined by a model parameter known as temperature. 
A higher value of temperature means that there are higher probabilities that less likely tokens are generated. 
For structured data-extraction tasks, working at a temperature value of 0 is typically the best, as this will lead to deterministic outputs with the most relevant information. 
It is important to keep in mind that most \glspl{llm} are so-called autoregressive models.\autocite{radford2018improving, radford2019language, touvron2023llama} This means that they generate the output using their previous output tokens as input. For example, the model in \Cref{fig:background_image} would generate one token based on the input, and the output token would then be added to that input. Using this longer input, the model can be queried again, producing another token. This process can be repeated a fixed number of times or until a specific end-of-sequence indicator is reached.  
The autoregressive nature is important because it leads to two major limitations. First, making predictions requires a full pass through the model for each output token, which can be slow. Second, errors multiply, i.e., if a token is generated incorrectly, it cannot be \enquote{fixed}, and the generation of subsequent tokens is affected by that error, potentially leading astray.\autocite{dziri2023faith}

\paragraph{Embeddings} Internally, the models do not work with the discrete tokens. Instead, they operate with high-dimensional vector representations of tokens, so-called embeddings. The power of these embeddings is that they capture the syntactic and semantic relationships between tokens. That is, synonyms or related words and phrases are mapped into neighboring parts of this high-dimensional space.
Internally, \glspl{llm} transform input embeddings by letting them interact, e.g., using the attention mechanism\autocite{vaswani2023attention} to give access to the entire sequence at each time step. The attention blocks are the core of recent \glspl{llm}, allowing them to capture long-range dependencies and making the training of these models highly parallelizable. However, attention does not account for changes in the order of the tokens. To still capture the order of the tokens, \glspl{llm} introduce so-called positional encoding that adds a unique signal to each word embedding, which describes the location of each token within the input.

\paragraph{Training and tuning of LLMs} The capacity of these models to generate coherent text comes from their training, in which they are exposed to massive amounts of natural text mined from the Internet. 
Most recent \glspl{llm} are initially trained to predict the next token within a sentence. This process is called pre-training and provides models with general-purpose representations, an understanding of semantics and syntax, and world knowledge.\autocite{zhou2023lima}
Since those models are trained to complete text, they are typically not ideal for answering questions. They will often attempt to complete input patterns, e.g., by generating more questions. 
To overcome this issue, they are often specifically tuned to follow desired guidelines and to answer such as humans would do. This is known as instruction-tuning and is typically performed as a supervised training task, where the model is trained on prompts and desired completions. In the context of \glspl{llm}, prompts refer to the input sequence containing the user's query and potentially some additional examples (see \Cref{sec:prompting}).
The most recent models are often also aligned to human preferences using processes such as \gls{rlhf} in which \glspl{llm} are tuned based on preference data such as the preferred choice among two completions generated by the model.\autocite{ouyang2022training} Importantly, it has been observed that this process makes the output probabilities of the model less calibrated,\autocite{lyu2024calibratinglargelanguagemodels} that is, they no longer correspond to expected error rates.

\paragraph{\gls{llm}-systems} 

For simplicity, we are going to write mostly \glspl{llm} even though most of the time, we are using systems. Systems are a combination of models and additional tools that might include safety filters. When a user calls a model via an \gls{api}, she seldom interacts with only the model but also has some inputs and outputs being processed by additional tools.

Additional resources on how \glspl{llm} work can be found in the \OnlineMaterial{Overview of the working principles of LLMs}{https://matextract.pub/content/background/resources_LLMs.html}.
\section{Structured data extraction workflow}
\label{sec:workflow}

\begin{figure}[htbp]
    \centering
    \includegraphics[width=\textwidth]{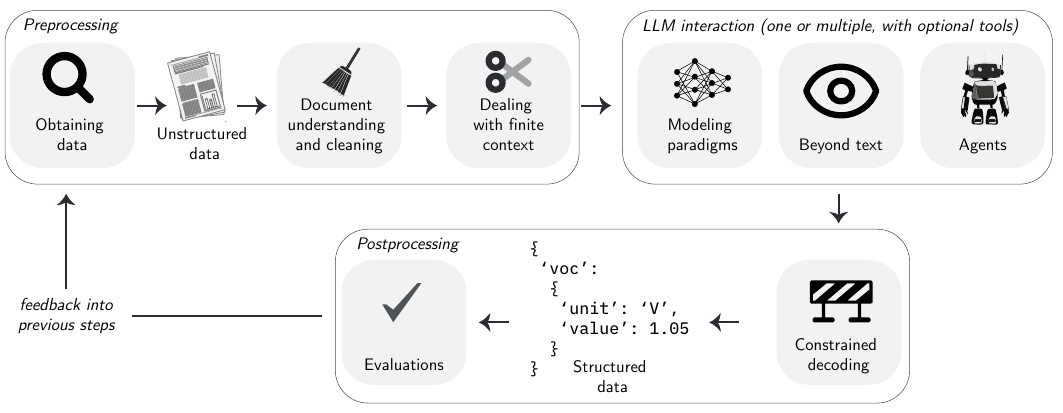}
    \caption{\textbf{Data extraction workflow}. This figure illustrates the flow of data from left to right through various stages of the extraction process. The evaluation loop includes all steps in the workflow, indicating that if evaluations do not yield satisfactory results, corrections and improvements may be necessary at any stage. It is important to conduct these evaluations using a representative and labeled test set, rather than the entire unstructured data corpus. Once the evaluations demonstrate satisfactory results, the entire corpus of unstructured data can be processed.}
    \label{fig:general_figure2}
\end{figure}

The structure of this section describes a common workflow of a data extraction project.
In practice, parts of the workflow must be optimized iteratively depending on the quality of the extraction result (see \Cref{fig:general_figure2}). The pipeline typically starts by collecting data (\Cref{sec:obtaining_data}), which then must be preprocessed and often chunked before it can be sent to interact with an \gls{llm}. The \gls{llm} can be used in diverse ways, such as by prompting or fine-tuning it, or in agentic (to use tools) or multimodal (to study other modalities than text) settings. Often, the desired output follows a special structure. In those cases, one can leverage techniques to ensure that the output follows this structure. 
To be able to optimize the extraction, it is essential to evaluate the extraction performance. This is often not trivial but can be aided by insights from materials science and chemistry. 
\subsection{Preprocessing}
\subsubsection{Obtaining data} \label{sec:obtaining_data}
The first step in the structured data extraction workflow is acquiring the necessary data. 
Data mining can pose significant challenges due to legal restrictions and practical difficulties in accessing and processing scientific literature. To ensure the legality of data mining, it is important to consider copyright infringements that may occur when copying and adapting the original work. There are two possibilities to avoid copyright infringements: either by getting the owners' permission for usage, or if the data mining actions fall under a copyright exception. Only a few publishers, such as Elsevier, Wiley, and Springer Nature, provide a general copyright license for \gls{tdm} use in addition to their usual contracts, and sometimes at an extra cost. 
However, most publishers lack a general \gls{tdm} agreement. Thus, researchers (or the libraries that have an agreement with the publishers) must approach the publishers individually.  
Yet, there is a wide variety of copyright exceptions that may also cover \gls{tdm} research. In the US, there is a \enquote{fair use} clause that represents a general exception to copyright for research and educational purposes, which has so far been interpreted by courts to allow at least some \gls{tdm} applications. \autocite{authorsguild2012} In contrast, the UK and Japan have explicit exceptions to copyright for content mining, and EU member states have to at least permit copying for non-commercial research or private studies.\autocite{Molloy_2016, FiilFlynn2022}

\begin{table}
    \centering
    \caption{\textbf{Overview of some data sources relevant for structured data extraction from scientific text, including published articles in open-access archives data dumps}. This table gives a non-comprehensive overview of data sources, the license provided by these publishers, the database size, the available data format of the articles, and how these articles can be accessed programmatically.}
    \begin{threeparttable}
    \footnotesize
    \begin{tabularx}{\textwidth}{p{4em}XXXXX}
        \toprule
        data source & description & license & size & data format & automated access \\ 
        \midrule 
        \multicolumn{4}{l}{\emph{Published articles}} \\
        \midrule
        EuroPMC \autocite{europepmc}&Corpus of life science literature&License terms of the respective authors or publishers &43.9 M abstracts, 9.7 M full text articles&XML or PDF& With RESTful \gls{api} from EuroPMC\\
        arXiv \autocite{arxiv}& OA\tnote{1} archive for different academic disciplines &License terms of the respective authors or publishers\tnote{2} &2.4 M full text articles &HTML or PDF&With arXiv \gls{api}\\
        ChemRxiv \autocite{chemrxiv}&OA\tnote{1} archive for Chemistry and related areas &License terms of the respective authors or publishers &150 K full text articles&PDF&With ChemRxiv Public API\\ 
        USPTO\tnote{3} &Federal agency for granting U.S. patents and registering trademarks& Text and drawings are typically not subject to copyright restrictions  &Over 50K chemical reactions only from 1976 till 2016 &Images and text documents&With USPTO APIs \\
        \midrule
        \multicolumn{4}{l}{\emph{Data dumps}} \\
        \midrule
        S2ORC \autocite{s2orc2019} &Academic articles from many academic disciplines &ODC-BY 1.0\tnote{4} &81.1 M abstracts, 8.1 M full text articles&Machine readable text&With Semantic Scholar APIs \\
        Elsevier OA CC-BY\tnote{1} \autocite{elsevieroaccby2020} & Corpus of OA\tnote{1} articles from across Elsevier’s journals &CC-BY\tnote{5} &40 K full text articles & &With Elsevier's APIs \\
        Open Reaction Database \autocite{opendb, Kearnes2021} &OA\tnote{1} corpus for organic reaction data & CC BY-SA\tnote{6} & 2 M chemical reactions & &With \href{https://github.com/open-reaction-database/ord-interface}{ORD interface} \\
        \bottomrule
    \end{tabularx}
    \begin{tablenotes}
            \item[1] OA is 'Open-access'.
            \item[2] For arXiv one has to check the copyright restrictions for each article. The default license does not specify reuse. 
            \item[3] USPTO is 'The United States Patent and Trademark Office'.
            \item[4] ODC-BY 1.0 is 'Open Data Commons Attribution License v1.0'.
            \item[5] CC BY is 'Creative Commons Attribution'.
            \item[6] CC BY-SA is 'Creative Commons Attribution - ShareAlike'.
        \end{tablenotes}
    \end{threeparttable}
    \label{tab:datasources}
    \normalsize
\end{table}

Several resources besides the major scientific publishers are available for data mining (\Cref{tab:datasources}). 
First, there are repositories of open-access published articles like EuroPMC\autocite{europepmc} or preprint servers such as ChemRxiv\autocite{chemrxiv} and arXiv\autocite{arxiv}, which contain a wealth of open-access scientific articles about different academic disciplines. 
In addition, there is the SciHub corpus,\autocite{scihub} which contains more than 88\,M also copyright-restricted articles. 
However, it might be illegal to use in certain jurisdictions.
Furthermore, data dumps like S2ORC\autocite{s2orc2019} or the Elsevier OA Corpus\autocite{elsevieroaccby2020}  collect a large variety of open-access scientific publications with mostly Creative Commons Attribution licenses. Most available data sources also provide automated access through an \gls{api}, simplifying data collection. However, most \glspl{api} provide a limit for requests per time to protect servers from overload. Thus, \gls{api} providers might block researchers' IP addresses or even cut off whole institutional access if \gls{tdm} rules are violated.\autocite{Baldi_2011, Molloy_2016}

\paragraph{Tools for data mining}
Various tools have been developed to aid in the data mining process. Thus, for the first step in this process, typically the collection of a database of potentially relevant articles for a topic, one could use different techniques like full-text or abstract search, search across different sources, so-called federated search, or search for one specific publisher. 
The Crossref \gls{api} \autocite{Lammey2015} or the Scopus \gls{api} \autocite{elsevierDevPortal2023} are examples of federated search \glspl{api}, which can be used to collect potentially relevant works from various publishers and sources and retrieve their metadata. However, it is important to note that the selection of relevant articles for a given research question is still an open research question. 
After finding relevant articles, one could use tools like \texttt{SciCrawler} \autocite{scicrawler2023} and \texttt{Scidownl} \autocite{scidownl2023} to collect the full-text articles from different corpuses, although researchers must be aware of current copyright law. Articles from open-access databases can be collected using tools such as \texttt{pygetpapers}.\autocite{pygetpapers2023} One can find a demonstration of these tools in the \OnlineMaterial{Obtaining data}{https://matextract.pub/content/obtaining_data/index.html}.

\paragraph{Importance of structured data}
However, simply mining unstructured data is insufficient for creating a data extraction pipeline. To optimize and evaluate the extraction pipeline, it is critical to programmatically evaluate the extraction performance at a later stage. This requires a set of unstructured data paired with the desired output. Tools such as Argilla\autocite{argilla2022} or Doccano\autocite{doccano} are emerging to aid in the creation of such datasets. Since extraction pipelines need to be optimized, such as by fine-tuning the \glspl{llm}, it is necessary to have both a test set and a validation set. The validation set is necessary because the test set should only be used to evaluate the final performance of the pipeline. If the test set is used to guide the optimization of the pipeline, information from the testing stage will influence the optimization process (data leakage), which is a major pitfall in \gls{ml}.\autocite{Kapoor_2023} 

\subsubsection{Curating and cleaning data}
\label{chap:parsing_cleaning}

\begin{figure}[!htb]
    \centering
    \includegraphics[]{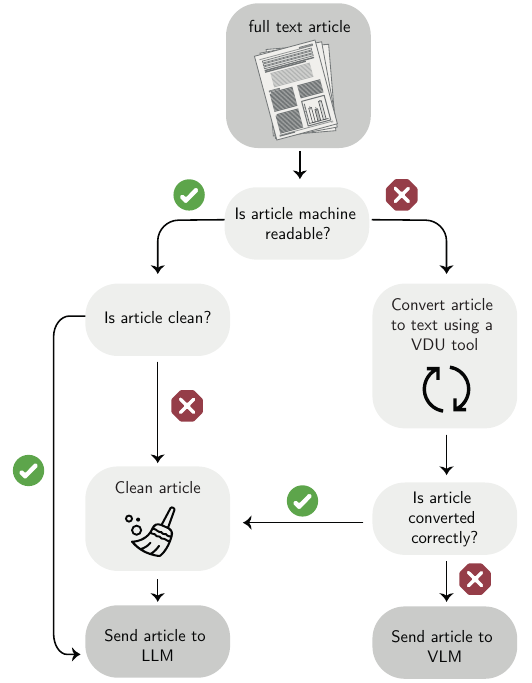}
    \caption{\textbf{Data preprocessing workflow.} The process from the mined articles to machine-readable and cleaned format, which one could send to a \gls{llm}. For articles for which the relevant information cannot readily be extracted using conventional \gls{vdu} tools, \gls{vlm} might be a suitable alternative (see \Cref{chap:multimodal_models}).}
    \label{fig:dt_document_parsing}
\end{figure}

While there are plenty of resources that provide (unstructured) data, this data is often not in a form that can, or should, be used for generative data extraction (see \Cref{fig:dt_document_parsing}). 
On the one hand, this is because the data might be only provided as an image, which cannot be used as input for a text-based model. 
On the other hand, the unedited articles often still contain information (e.g., bibliography, acknowledgments, page numbers) that might not be needed for the extraction tasks. 
Thus, the input for the model could be compressed and less \enquote{distracting} if such parts are removed.

\paragraph{Document parsing and understanding}
Before any such curation can occur, the document must be in a form that curation tools and models can work with. In many cases, this involves a step of \gls{vdu}, which enables machines to understand documents. 
This might require not only \gls{ocr}, which is focused on extracting machine-readable text from text images, but also the ordering and sequence of text blocks and layout (i.e., the reading order). Further layout understanding is needed because documents such as images or PDFs, in contrast to \gls{xml} or \gls{html}, do not have semantic annotations that describe the meaning and context of different parts of the document.
In such cases, additional tools must be used to analyze the document's structure. 
Traditionally, this has often been performed using rule-based systems (as, e.g., in PDFDataExtractor by \textcite{Zhu_2022}) operating on \gls{ocr} output (which can be obtained using \gls{ocr} tools such as Tesseract\autocite{smith2007overview}). 
However, recently, it has been shown that more general higher-performance approaches can be obtained using end-to-end modeling approaches such as \nougat\autocite{blecher2023nougat} or \marker\autocite{marker2023}, which are specialized for converting scientific papers from PDF to Markdown (see \OnlineMaterial{Document parsing with OCR tools}{https://matextract.pub/content/document_parsing_and_cleaning/parsing.html}). In those Markdown files, the meaning of different parts (e.g., sections, tables, equations) is semantically annotated using the Markdown markup.  Complicated structures like complex equations and large tables have a high potential for errors. \autocite{Meuschke2023} As an alternative to text-based models, multi-modal models can also handle other types of inputs, e.g., images. These approaches are discussed in \Cref{chap:multimodal_models}.

\paragraph{Document cleaning}
For many data extraction tasks, it can be useful to preprocess the data, for example, by deleting text that does not contain the relevant information. This can make the system more robust (as noise is removed) and cheaper (as less text has to be processed).  
Often, sections and structures such as authors, affiliations, acknowledgments, references, page numbers, headers, and footers, as well as unneeded white spaces or linebreaks, can be removed. Some old articles include the end or the beginning of the next article in the text file, which could lead to major errors during the extraction. For chemistry and materials science-related data extraction tasks, for example, one might need only the experimental part to extract raw materials, synthesis, characterization, and products, or one need to include the Supporting Information files, which usually contain detailed descriptions of synthetic routes and experimental procedures in organic chemistry.\autocite{bran2024knowledge} Hence, the content length of the articles could be reduced to the most relevant parts.  
To do so, practitioners often use regular expression-based pipelines as, for example, in the ChemNLP project.\autocite{chemnlp2023} One can find examples for data cleaning methods in the \OnlineMaterial{Document cleaning}{https://matextract.pub/content/document_parsing_and_cleaning/cleaning.html}.

\subsubsection{Dealing with finite context} \label{chunking}
\begin{figure}[htbp]
    \centering
    \includegraphics[width=1\textwidth]{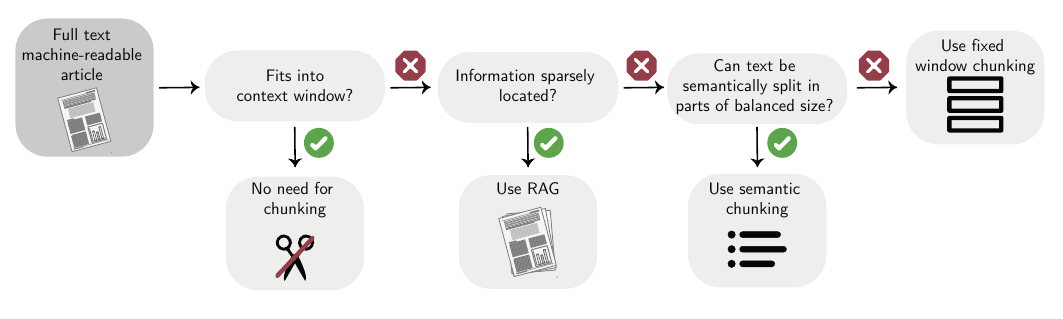}
    \caption{\textbf{Decision tree to help decide what chunking strategy to use.} If the input text is quite short, no chunking is required. In contrast, if the information is spread across a very large corpus, \gls{rag} can provide cost and efficiency benefits. Chunking is typically applied with \gls{rag}. In this case, one can use semantic chunking if it provides chunks that fit into the context window. The most simple option is to chunk text using a fixed window size.}
    \label{fig:chunking_decision_tree}
\end{figure}

The context window is the range of tokens a model can process at a time. 
This includes all text, such as the input given to the \gls{llm} and the response it produces. When using an \gls{llm} with long, low information-density documents, the context window size can be a hurdle.
In such cases, chunking strategies can be used to fit the text in pieces (\enquote{chunks}) into the context window. However, careful consideration must be given when chunking documents, as the entity of interest could be located at any position in the text.\autocite{buehler2024accelerating}
This consideration is particularly critical in chemical sciences literature, where key information is often dispersed throughout the document. For instance, abbreviations and nomenclature are typically introduced at the beginning, while synthesis procedures and experimental results may be described in separate sections.
Therefore, the choice of chunking strategy is crucial (see \Cref{fig:chunking_decision_tree}).
Importantly, even though the context windows of \glspl{llm} are rapidly growing,\autocite{naveed2024comprehensiveoverviewlargelanguage} there is also value in chunking because models can perform better if given a higher density of relevant content.\autocite{skarlinski2024languageagentsachievesuperhuman}

\paragraph{Chunking techniques} The simplest approach is to divide the text into chunks of identical size to fit in the context window. This is also known as fixed-size chunking. One downside of this approach is that a word could be split between two chunks. One can improve upon this by splitting based on special characters like periods (\texttt{.}) or newlines (\texttt{\textbackslash n}). In most situations, this is similar to splitting based on sentences and will hence leave words intact, but it is problematic when the text, such as scientific text, has numbers with decimal points. For our use case, it is important not to split these numbers.

Even though this approach preserves sentences, chunks of text are often related to the previous chunk. To help preserve some more semantic context from neighboring chunks, one can have a certain length of overlap between chunks. This, in practice, has been shown to produce better results.\autocite{carta2023iterativezeroshotllmprompting} The size of this overlap plays an important role in contextual awareness and the total number of required \gls{llm} queries.

In many cases, we can also use the fact that many documents are naturally split with headings and sections. We can use this to our advantage on top of all chunking techniques. This is called \emph{semantic chunking}.

\paragraph{Using RAG and classification to save costs}
If the information density in the documents is low and one has to deal with many documents, one can store these chunks in a database and retrieve relevant chunks when needed to query the \gls{llm}. Typically, a vector database is used to store embeddings of the chunks while enabling fast retrieval of semantically relevant chunks. This process of retrieving information and using it as additional context when querying an \gls{llm} is called \gls{rag}.\autocite{lewis2021retrievalaugmented}
Since embedding text is cheaper than running it through a generative \gls{llm}, this can be an effective strategy when dealing with many documents. 
However, the performance of this technique highly depends on the embedding model and retrieval strategy used.

A classification model can also be used to determine a text's relevance to the tasks at hand to improve extraction efficiency by, for example, determining prior to extraction whether a chunk contains relevant information.\autocite{Zheng_2023, Choi2024} The classification models used for this task can range from logistic regression trained on the text embeddings to using the \gls{llm} directly to classify text. 

Chuncking techniques and \gls{rag} are demonstrated in \OnlineMaterial{Strategies to tackle context window limitations}{https://matextract.pub/content/context_window/Dealing_with_context_window.html}.

\paragraph{Extending context windows} 
Attention, which is a key part of most \glspl{llm}, is costly to scale because the size of the attention matrix grows quadratically with the sequence length. A number of techniques have been developed to increase the model's context window.

\subparagraph{Architectural changes (prior to training)}
One approach is to modify the model architecture so that it can manage larger context windows without a corresponding increase in computational cost. This can either be done by giving the model access to the previous hidden state (but not computing the gradients for the old states)\autocite{dai2019transformerxl} or by using an additional (external) memory. This can be, for example, in the form of a small number of extra tokens that store \enquote{global state}\autocite{beltagy2020longformer} or in a global memory to which context is added during training and which can be retrieved using elements of the conventional attention mechanism. Those techniques currently still share a problem that was also faced by \glspl{rnn}: adding new information to a fixed-size memory will dilute past memory over time.

\subparagraph{Changes after training}
In addition to changes to the actual model architecture, some changes can also be performed after training. For instance, models trained with so-called \gls{rope}\autocite{su2023roformer}, such as \llamathree can be fine-tuned to understand (relative) position of tokens over a range larger than the context window used in the actual training.\autocite{chen2023extending} While this does not increase the actual context window, it can provide the model with a better understanding of the position of a chunk within a larger document.

\subsection{LLM-Interaction}

\begin{figure}[htbp]
    \centering
    \includegraphics[width=1\textwidth]{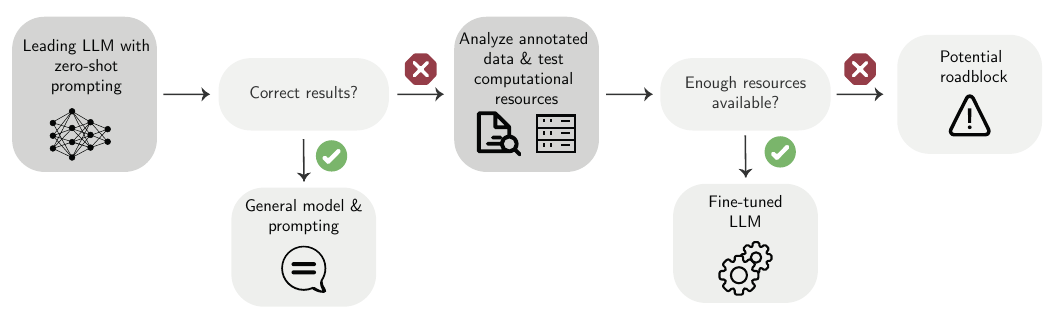}
    \caption{\textbf{Decision tree on how to decide which learning paradigm is better for each case}. The term \enquote{leading \gls{llm}} is intended to describe any general purpose \gls{llm}. The decision normally starts by testing a leading \gls{llm} with zero- or few-shot prompts. If the results are good enough, it is possible to continue with this method; if not, fine-tuning may be necessary. But fine-tuning requires additional labeled data and computational resources. If those are missing and simpler approaches do not work, pre-training also does not provide a solution, as it typically requires even more data (even though unlabeled) and computational resources.}\label{fig:fine_tuning_image}

\end{figure}

Once the data from different sources is correctly parsed, the next step in the extraction workflow is to select the right modeling framework to proceed. This section aims to provide insights about the different models and learning paradigms. Furthermore, it presents the advantages, challenges, and potential strategies for optimizing the utilization of these tools in extracting structured data.

\subsubsection{General model with prompt engineering} \label{sec:prompting}

In principle, as pointed out in \Cref{fig:fine_tuning_image}, using a leading general-purpose \gls{llm}\autocite{chiang2024chatbot, mirza2024large, xie_creation_2024, laurent2024labbenchmeasuringcapabilitieslanguage} may seem a good first option due to its simplicity since no further training is needed. 
This option allows for rapid deployment and does not need further labeled data beyond the data needed for the evaluation of the extraction task. Such base models have been trained on a general corpus of data that does not belong to any specific field in particular and has mostly been mined from the Internet. Thus, these models may lack knowledge and understanding of field-specific vocabulary.\autocite{udandarao2024zeroshot, CastroNascimento2023, White2023_assessment, xie2023darwin, hira2024reconstructing} However, as a result of the amount of data they were trained on (e.g., the new \llamathree model has been trained on over 15\,T tokens of publicly available data\autocite{Meta_Llama_2024, the_Llama_3_Herd_of_Models}), these models are often able to yield acceptable results for any of the typical \gls{nlp} tasks for data extraction, such as \gls{ner} or \gls{re} without further tuning.\autocite{hu2024improving} It is important to note, however, that it is always valuable to try multiple models as large models might be better at recall but show a lower precision as, for example, \textcite{bran2024knowledge} found for knowledge graph extraction from total synthesis documents.

When applying \glspl{llm} to extract information, one can adapt the model to the task by changing the prompt. Besides the actual text from which structured data should be extracted, the prompt contains the instructions that are given to the model as guidance to obtain the desired results. 
In addition, some advanced prompting techniques such as \gls{cot}\autocite{wei2023chainofthought} have been shown to improve performance on various tasks. However, simple so-called zero-shot or few-shot prompting approaches are still the most common for generative data extraction. 

\paragraph{Zero-shot prompting}
Zero-shot prompting means that the text which is provided to the model does not contain any examples or demonstrations; it will only contain the task that the \gls{llm} is supposed to do (as well as the input text). When using this zero-shot prompt, the results solely depend on the knowledge and reasoning capabilities the model acquired during its training. If the model is unfamiliar with the task and vocabulary of the field, the outcomes are unlikely to meet the expected level of satisfaction, i.e., the \glspl{llm} might not recognize dimethylformamide (DMF) as a solvent in synthesis or might not correctly understand how the compositions are expressed\autocite{hira2024reconstructing}. 

Despite the simplicity, encouraging results have been obtained in this setting.\autocite{Zheng_2023, kommineni2024humanexpertsmachinesllm}
An alternative is to include some advanced prompting techniques. In this fashion, Polak and Morgan\autocite{Polak_2024} prompted \gptthreefive, \gptfour, and \llamatwo models to extract information about materials such as pooling rates of metallic glasses and yield strengths of high entropy alloys. To make their zero-shot prompt more reliable, they developed a workflow with several follow-up questions. Within these follow-up questions, they included a technique called self-reflection,\autocite{shinn2023reflexion} which prompts the model to check its answers, looking for possible errors and correcting them if they exist. 

\paragraph{Few-shot prompting}
One of the most fascinating observations made in the development of \gptthree is that \glspl{llm} can perform so-called in-context learning,\autocite{brown2020language} i.e., learn a new concept without updating the model parameters but by only showing examples of desired behavior in the prompt.\autocite{kaplan2020scaling} Sometimes, even only one example (one-shot learning) is enough.\autocite{pmlrv225goel23a} However, often more  ($k$) examples (few-shot learning) lead to better performance, mainly because adding more examples helps prevent the model from becoming too focused on any single example.\autocite{xu2023unleash}
The optimal number and order of examples will depend on each application highly and can be found by evaluating prompts with a variable number of shots.\autocite{lu2022fantastically} Often, as in the case of \textcite{Zhang2024chemicaltext} for chemical data extraction, more examples improve the performance, but the number of examples one can provide is limited by the context window of the model.
The examples that are provided must be chosen wisely, and it can be important to include not only positive examples but also negative ones, e.g., JSON schemas with some empty optional fields.\autocite{agrawal2022large} To choose the examples for the prompt, one approach proposed by \textcite{liu2021makes} is to use $k$-\gls{nn} clustering in the embedding space and build the prompt by including the $k$-neighbors for each test sample. Additionally, the examples can be modified to reflect the intended completion of the model better or to improve the conditions, e.g., summarize the examples, so fewer tokens are consumed (see \OnlineMaterial{Collecting data on the synthesis procedures of bio-based adsorbents}{https://matextract.pub/content/biomass_case/biomass_case.html}).

Examples of zero-shot and one-shot prompts are shown in \OnlineMaterial{Choosing the learning paradigm}{https://matextract.pub/content/finetune/choosing_paradigm.html}.

\paragraph{Advanced prompting techniques} Another possibility is to combine the few-shot prompt with other prompting techniques; for example, \textcite{sui2024table} combined the few-shot prompt with so-called self-augmentation. Self-augmentation consists of a two-step process in which the model is prompted twice, and the first answer of the model is used to improve the second prompt by giving insight into the data. Using this technique to prompt \gptthreefive and \gptfour models improved both models' results with respect to a simple one-shot prompt in better understanding and extracting data from tables of different datasets. They also observed that the performance increases significantly when going from zero-shot to one-shot prompts. One drawback that especially limits few-shot prompting is the context length of the models.\autocite{xie_creation_2024} 
Nevertheless, the latest models released notably improved the context window (e.g., the latest \geminipro model has a 2M token context window), allowing for a bigger number of shots\autocite{agarwal2024manyshot} (for further discussion about the context length, see \Cref{chunking}).
Despite all the good results that these advanced techniques can give, it is not always clear that they are going to work (see \OnlineMaterial{Collecting data on the synthesis procedures of bio-based adsorbents}{https://matextract.pub/content/biomass_case/biomass_case.html}).\autocite{stechly2024chainthoughtlessnessanalysiscot, ridnik2024codegenerationalphacodiumprompt} However, due mainly to their ease of use, it is worth trying them before moving on to more complicated setups.

Many recent models differentiate between so-called system and user prompts. The system prompt provides high-level instructions or context that guides the model's behavior (\enquote{persona}). It sets the overall tone or rules for the conversation.
However, the benefits of role prompting are still being debated.\autocite{White2023_assessment}
A simple example of a system prompt for the data extraction task can be: \enquote{You are a chemistry expert assistant and your task is to extract certain information from a given text fragment. If no information is provided for some variables, return \texttt{NULL}}. The system prompt is typically sent with the first completion and cannot be overridden by subsequent user follow-up questions. On the other hand, a user prompt represents the actual queries, statements, or prompts given by the user. They might contain the input text from which one wants to extract data, as well as a few-shot examples.

\paragraph{Prompt optimization} Typically, prompting techniques are implemented by hard-coding templates. This task can be made easier using frameworks like LangChain\autocite{Chase_LangChain} or LlamaIndex\autocite{Liu_LlamaIndex_2022}, which provide some of those templates. 
However, the models have varying preferences for prompts,\autocite{sclar2023quantifying} making actual prompting mostly empirical (or often manual and prone to data leakage) and constantly changing. This makes it impossible to create general and robust prompts.
To address this, frameworks such as DSPy have been developed.\autocite{khattab2023dspy} These systems perform the optimization of entire \gls{llm} pipelines automatically and systematically (e.g., by autogenerating prompts or few-shot examples and then choosing the best ones using cross-validation). However, those approaches are still in their infancy and often require many \gls{llm} calls to optimize prompts.

\paragraph{Schema format} Another problem when using \glspl{llm} is the cost associated with their use. While trying to improve the performance when prompting the \gptthreefive model for chemical data extraction, \textcite{patiny_automatic_2023} reduced the number of tokens by using a YAML input schema instead of the common JSON format.  
Additionally, they proved that the \gptthreefive model can perform well with YAML schemas. However, the performance of \glspl{llm} when working with different types of formats will vary among models depending on the training data\autocite{xia2024fofobenchmarkevaluatellms, sui2024table} (see \OnlineMaterial{Data annotation}{https://matextract.pub/content/obtaining_data/annotation.html} for examples).

\begin{table}
    \centering
       \caption{\textbf{Key factors to consider for each learning paradigm assuming that the model is already deployed, either by API or by other means.} Note that we assume a first-year chemistry student is the user and consider the use cases in which the student may be interested. We assume for all cases the use of proper \glspl{llm} (\textgreater1\,B parameters). 
       \enquote{Prototyping time} key factor refers to the time that it takes the user to do the first tests and see if the model can be used for that task. 
       \enquote{Field-specificity} intent to explain the understanding that the model can deal with terminology that is particular to a certain field, such as chemical formulas. This factor is especially related to the tokenizers of the models. Tokenizers need to be set prior to training. Hence, tokens relevant to a particular topic might be missing from pre-trained tokenizers. In pre-training or fine-tuning additional tokens can be added, which can result in an improvement when extracting data from texts with complex notation.}
    \begin{threeparttable}
 
    \footnotesize
    \begin{tabularx}{\textwidth}{p{7em}XXp{7em}p{7em}}
        \toprule
        Key factors & Zero-shot prompting & Few-shot prompting & Fine-tuning & Pre-training \\ 
        \midrule
            Expertise & Very low & Very low & Medium/High\tnote{1} & High \\
            Prototyping time & Minutes & Hours & Days & Weeks \\
            Field-specificity & None\tnote{2} & Low\tnote{2} & High & Very high \\
            Task-specificity & None\tnote{2} & Low\tnote{2} & High & Medium \\
            Data & None & $10^0$-$10^1$ examples\tnote{3} & $10^2$-$10^3$ samples & $\approx10^{12}$ tokens \\
        \bottomrule
    \end{tabularx}
    \begin{tablenotes}
            \item[1] Depending on how much data is available or needs to be cleaned and prepared.
            \item[2] Considering knowledge added to the general pre-training of the model during inference.
            \item[3] The number stands for the traditional use of the few-shot paradigm. Thus, with the actual context length of some current models, the number of shots used can increase a lot.
        \end{tablenotes}
    \end{threeparttable}
    \label{tab:pros_and_cons_paradigms}
    \normalsize
\end{table}

\subsubsection{Fine-tuning}
When the leading models fail to produce satisfactory results, fine-tuning can be used to improve a model for a particular task or domain by training it further on smaller but well-curated task-specific datasets.\autocite{shamsabadi2024large,Luu_2023,van2024assessment,Kim_2024,jablonka2024leveraging}  It is important to point out that fine-tuning an \gls{llm} still requires a lot of expertise, which can be an important drawback (\Cref{tab:pros_and_cons_paradigms}). However, \textcite{Zhang2024chemicaltext} found that by fine-tuning leading models (including \gptthreefive) on chemical data extraction tasks (such as reaction role labeling, metal-organic framework (MOF) synthesis information extraction, NMR data extraction) they could often outperform bespoke domain-specific models.

An example of how to fine-tune a model using an existing dataset is available at \OnlineMaterial{Choosing the learning paradigm}{https://matextract.pub/content/finetune/choosing_paradigm.html}.

\paragraph{Fine-tuning techniques}  In principle, one can perform fine-tuning by changing all trainable weights of the model (\enquote{full fine-tuning}). This, however, results in a very computationally demanding and slow process. To address this, \gls{peft} techniques, in which only a few parameters (frequently only \SI{1}{\percent} of the total number of trainable parameters) are updated, have been developed.
One popular \gls{peft} technique is called \gls{lora}.\autocite{hu2021lora} \Gls{lora} involves freezing the weights and then decomposing the updates' matrix (a matrix that contains the weight updates during the fine-tuning) into two lower-rank matrices that can be optimized during fine-tuning. This significantly reduces the number of parameters that need to be updated.
Often, fine-tuning is done in a task-specific way, for example, focusing on \gls{ner}+\gls{re}\autocite{dagdelen_structured_2024}. For instance, \textcite{dagdelen_structured_2024} fine-tuned models to link dopants and host materials, and for the extraction of general information (chemical formulae, applications, and further descriptions) about metal-organic frameworks and inorganic materials.
Fine-tuned \glspl{llm} have also been used for constructing a materials knowledge graph\autocite{ye_construction_2024}, and extracting nanoparticle synthesis procedures.\autocite{Lee_2024_synthesis}
Given enough diverse data, it is also possible to fine-tune models for general \gls{ie} applications.  For example, \textcite{sainz2024gollie} built a model called GoLLIE to follow task-specific guidelines, improving zero-shot results on unseen \gls{ie} tasks.

One important drawback of fine-tuning is that it modifies the pre-trained model, potentially limiting its generalizability. This has been observed by \textcite{Ai_2024}, who used the Llama-Adapter\autocite{zhang2023llamaadapter} method (another \gls{peft} technique) to fine-tune a \llamatwoseven model for the extraction of reaction information from organic synthesis procedures.
Interestingly, it has been shown that models tuned with \gls{lora} are less prone to forget the general capabilities of the base model but are also less proficient in learning new capabilities compared to full fine-tuning.\autocite{ivison2023camels, biderman2024lora} 

\paragraph{Human-in-the-loop annotation} One of the main limitations of fine-tuning is the need for large amounts of annotated data. 
To solve this problem, \textcite{dagdelen_structured_2024} proposed a human-in-the-loop annotation process in which a model performs the initial data annotation. A human expert corrects possible errors in the annotated data, and the resulting corrected data is used to further fine-tune the model that continues to label the data. Using this annotation method, they reduced the time needed to annotate each sample by more than half for the last annotated abstracts since correction is faster than annotating from scratch.

\subsubsection{Pre-training}
If fine-tuning falls short, one costly resort might be to train a model from scratch. For \glspl{llm} this usually requires prohibitory computational resources, datasets, and expertise, as described in \Cref{tab:pros_and_cons_paradigms}. For example, training the new \llamathree models needed almost 8\,M GPU hours using two clusters with 24\,K GPU each.\autocite{Meta_Llama_2024, the_Llama_3_Herd_of_Models} However, it is possible to fine-tune the \llamathreeseventy model using a single GPU with 80\,GB of memory.\autocite{dagdelen_structured_2024}

However, when the data to extract is very specific and general capabilities are not required, good results can be achieved using smaller \gls{lm} such as the ones of the BERT series (so-called encoder-only \glspl{lm}).\autocite{Zhang2022} For those smaller models, less training data and less computational resources are required. For example, SciBERT was pre-trained on a corpus of 1.14\,M articles, resulting in 3.17\,B words.\autocite{beltagy2019scibert} Similarly, MatSciBERT was trained on 3.45\,B words (compared to 300\,B tokens for \gptthree).\autocite{Gupta2022_matscibert} 
Note that for those smaller models, it is still recommended to fine-tune on the task that the model is intended to perform. In this way, Cole et al.\ built task-specific BERT models by pre-training them on batteries,\autocite{Huang_2022} optical materials,\autocite{Zhao2023} or photocatalytic water splitting\autocite{Isazawa2024} field-specific text extracted from scientific articles. After the pre-training, they fine-tuned them on the Q\&A task using the Stanford question-answering (SQuAD) v2.0 dataset.\autocite{rajpurkar2018know}
This pre-training and fine-tuning approach of smaller encoder-only models has also been applied to extract polymer property data\autocite{Shetty2023} and synthesis protocols of heterogeneous catalysts.\autocite{Suvarna2023}

Even though training smaller \glspl{lm} needs less data and computation compared to training an \glspl{llm}, training a small BERT from scratch might be more costly than performing fine-tuning using \gls{lora}. As shown by \textcite{song-etal-2023-honeybee}, \gls{lora}-based instruction fine-tuning of Llama models can outperform BERT-based models on a suite of materials science tasks relevant for data extraction.\autocite{song-etal-2023-matsci} Thus, before deciding to train an \gls{llm} for specialized tasks related to data extraction, we recommend exploring other options outlined in this review to find potentially feasible alternatives. 

\subsubsection{Beyond text}\label{chap:multimodal_models}
\begin{figure}[htbp]
    \centering
    \includegraphics[width=1\textwidth]{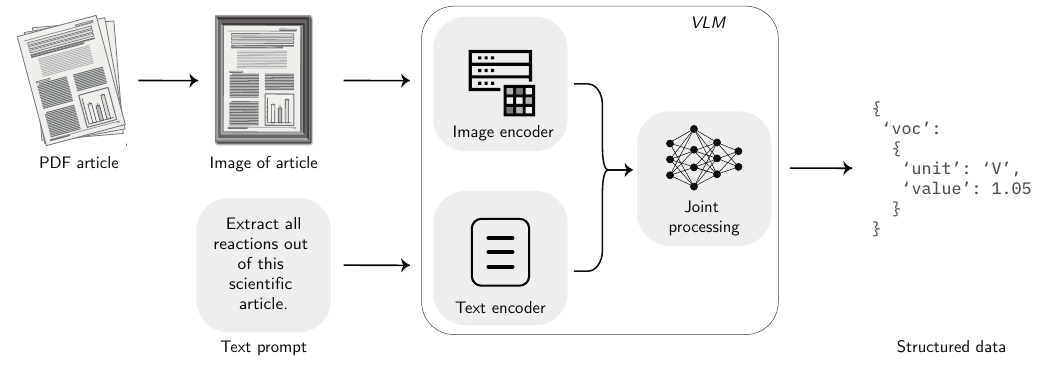}
    \caption{\textbf{Workflow of data extraction with \glspl{vlm}.} Papers, for example, in the form of PDFs, can be converted into images and then processed using an image encoder. The prompt containing the instruction is still provided in the form of text and the output of the \glspl{vlm} is structured data in the form of text.}\label{fig:VLM}
\end{figure}

\paragraph{Vision models}
While there are plenty of tools for converting text, e.g., in PDF documents, in machine-readable Markup language as discussed in \Cref{chap:parsing_cleaning}, the challenge of analyzing and converting complex structures like large tables or plots can often not be easily addressed with those tools. \autocite{circi2024using} This problem is becoming even more prominent in the extraction of chemical data, where, in addition to plots and tables, crystal structures, reaction schema, complex spectra, and intricate graphical representations contain vast amounts of critical information. 
Besides classic text processing models, there are also models that were specifically trained to understand and analyze images alongside text, which are so-called \glspl{vlm}. 
These models typically consist of three main components: an image encoder, a text encoder, and a component that fuses the information of these two.\autocite{weng2022vlm} 
The main advantages of the usage of such \glspl{vlm} are the additional information gained through images and plots and the end-to-end procession of the data since the loss of information due to conversion steps mentioned in \Cref{chap:parsing_cleaning} is minimized.\autocite{Zheng2024} Moreover, the end-to-end approach is preferred since models could choose and learn the best preprocessing method instead of humans choosing and hard-coding bespoke processing workflows. 
For instance, some of these models can leverage the specific layout of scientific articles, such as figures, tables, and captions, and, therefore, can extract the data more accurately.
The downsides of using \glspl{vlm} instead of \glspl{llm} are potentially higher costs for the data extraction. 
However, also \glspl{vlm} still requires some preprocessing of the input, like converting the input into suitable images and potentially removing irrelevant sections like acknowledgments or references or resizing and scaling the images. An example of a basic extraction workflow as shown at \Cref{fig:VLM} can be found in the \OnlineMaterial{Beyond text}{https://matextract.pub/content/beyond_text/beyond_images.html}.

A question that arises is in which case one should use an \gls{ocr}-\glspl{llm} extraction pipeline or a \gls{vlm} for extraction instead. 
Since usually \glspl{llm} model calls are less expensive, starting by testing these models instead of \glspl{vlm} is typically preferred for the extraction of mostly text-based articles.
However, if the input data includes a lot of complex structures like tables and plots,  a \gls{vlm} can often be a suitable choice, as demonstrated in the extraction of copolymerization reactions\autocite{schilling_Wilhelmi2024polymers} or electrosynthesis reactions\autocite{Leong_2024}---in particular if the focus is on obtaining results without building complex preprocessing or agentic pipelines.
For example, \textcite{lei_materials_2024} demonstrated the higher efficiency of a \gls{vlm} in the task of detection of the material present in micrographs than a text-only \gls{llm}.

A wide variety of open-source and commercial vision models are available. The \gptfourvision and \claudethree models are the largest and most widely used models. Although these are commercial, some open-source \gls{vlm} such as DeepSeekVL also have achieved good results.\autocite{lu2024deepseekvl} 
\textcite{Liu2024ocr} used \glspl{vlm} to perform different \gls{ocr} tasks and introduced a benchmark to compare the performance of different \glspl{vlm}. While for simple tasks like recognizing regular text, the \glspl{vlm} achieved performance comparable to state-of-the-art \gls{ocr}-tools, they performed poorly on unclear images, handwritten text, and adhering to task instructions, as well as in extracting knowledge graphs of organic chemistry reactions.\autocite{bran2024knowledge} \textcite{alampara2024probinglimitationsmultimodallanguage} have shown that leading \glspl{vlm} struggle to analyze basic spectra and extract basic from reaction schema and tables. They have also exposed that the current models still fail to understand the images' content. Therefore, using specific tools developed for these tasks might still notably increase the accuracy.

\paragraph{Tools for special inputs/modalities}
Besides these general vision models, several tools are available for specific tasks and structures, such as tables, chemical structures, and plots. 
Since tables in scientific articles can be quite complex, a dedicated focus on these could help to increase accuracy.\autocite{llamaindex2023, Lee_2024_oxidation} For instance, tools like TableTransformer\autocite{smock2021tabletransformer} or DISCOMAT\autocite{gupta2024discomatdistantlysupervisedcomposition} are specifically optimized for dealing with data in tables and \textcite{liu2023deplotoneshotvisuallanguage} reported a plot-to-table translation model.

In chemistry and materials science, the most important information about chemical components and reactions is often hidden in images of structural formulas and reaction schemes. Therefore, these non-machine-readable depictions have to be extracted separately from the text. For instance, the ReactionDataExtractor 2.0 \autocite{Wilary2023} opens the possibility of extracting a complete reaction scheme, including components and reaction conditions. OpenChemIE also extracts reaction data from text or figures.\autocite{Fan2024} Nevertheless, many of these tools encounter problems with variable end groups mostly noted as 'R-group'.\autocite{Rajan2020}

Another important modality for data extraction is plots and images. The WebPlotDigitizer V4 is an open-source tool to extract data from plot images to numerical data,\autocite{webplotdigitizer2023} as shown by \textcite{Zaki2022} to extract glasses information from graphs. In addition, there has been a focus on extracting data (e.g., shape, size, and distribution of the pictured particles) from microscopy images.\autocite{Mukaddem2019, vonChamier2021, Stuckner2022}

Due to the wealth of tools available and the various possible use cases (leading to a combinatorial explosion of possible combinations of tools), one can consider using an agentic approach described in \Cref{chap:agents} to automatize the usage of these. 

\subsubsection{Agentic Approaches}\label{chap:agents}

\begin{figure}[htbp]
    \centering
    \includegraphics[width=\textwidth]{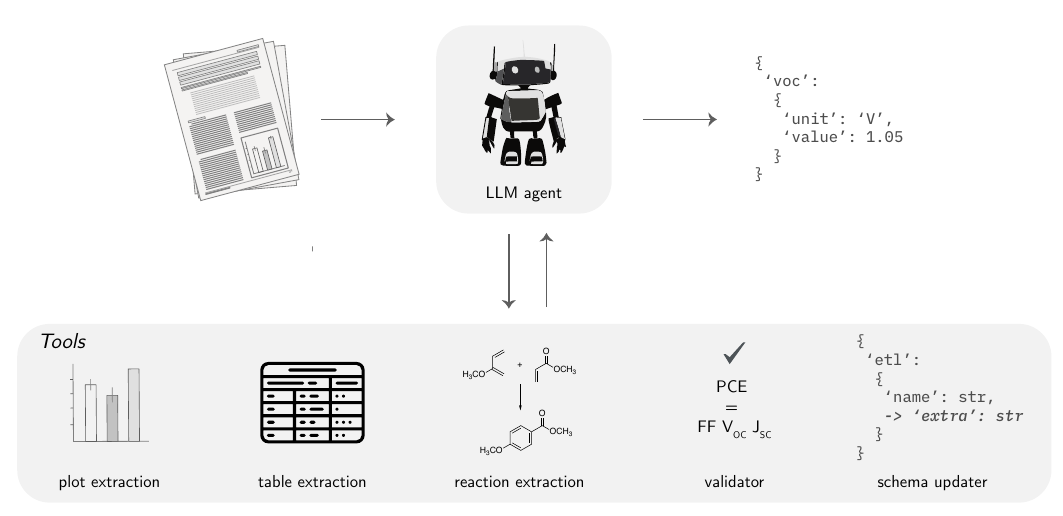}
    \caption{\textbf{General workflow of an agent for the data extraction task.} The unstructured data from the articles is given to the agent as text, images, equations in \LaTeX\ form, or any other format that can be given to the models. Using its reasoning capabilities, the agent decides which one of the available tools is best to extract each data type. When all the data available are extracted, the agent provides it as structured data.}
    \label{fig:agent_section}
\end{figure}

\gls{llm}-based agents are novel decision-making agents powered by one or more \gls{llm} that can interact with (and modify) an external environment.\autocite{gao2024empowering, weng2023agent}
They have shown promise in tasks as varied as simulating human behavior \autocite{park2023generative}, playing Minecraft \autocite{wang2023voyager}, molecule design,\autocite{liu2024teamaimadescientistsscientific, ghafarollahi2024protagentsproteindiscoverylarge, bou2024acegenreinforcementlearninggenerative} and autonomously performing chemical reactions.\autocite{MBran2024, Boiko2023}
As a result of the several reasoning cycles that the agents perform, they often outperform vanilla \glspl{llm}.\autocite{masterman2024landscape, xi2023risepotentiallargelanguage, ramos2024reviewlargelanguagemodels} But what advantages can they present in extracting data?

\paragraph{Why to use agents for structured data extraction?}

\subparagraph{Flexible dynamic workflows}
The use of agents can be especially interesting for data extraction when working with multimodal data, i.e., not only text but also tables and figures (\Cref{fig:agent_section}). 
Although \glspl{vlm} can understand figures, the results when extracting data might be not good enough\autocite{Liu2024ocr}, especially when this data is field-specific, such as specialized scientific images.\autocite{liu2023visual} 
Even though specialized tools exist for many of these use cases a limitation of all these specialized tools is that they must be manually used or manually chained into workflows by developing bespoke programs.
However, an \gls{llm} agent with access to these tools can autonomously build dynamic workflows without manual human input, making data extraction more accessible, scalable, and flexible.\autocite{ansari2023agentbased} 

\subparagraph{Higher accuracy}
Another important advantage of using agents is that they can show the ability to automatically improve and correct wrong results by using self-reflection and self-criticism of previous actions.\autocite{du2023improving} 
This ability has been shown to reduce hallucination considerably in some data extraction applications,\autocite{lala2023paperqa} even achieving superhuman performance.\autocite{skarlinski2024languageagentsachievesuperhuman} 

\paragraph{Design patterns for \gls{llm}-powered agents}

Different classifications have been proposed in the literature to define the agents.\autocite{xi2023risepotentiallargelanguage, gao2024empowering, Wang2024, sumers2024cognitivearchitectureslanguageagents} However, we think that the most appropriate for the data extraction task is a classification closer to the one proposed by \textcite{weng2023agent}.

\subparagraph{Planning}
Complex tasks typically involve many steps. Thus, agents need to be able to decompose tasks and plan ahead.
In \gls{llm}-agents, planning is typically provided by the reasoning capabilities of the \gls{llm}. The simplest form of task decomposition is \gls{cot} prompting, which utilizes test-time computation to decompose the task into simpler steps by prompting the model to \enquote{think step by step}. This framework has been extended in a multitude of ways, such as Tree of Thoughts,\autocite{yao2023tree} which, similar to \gls{cot} decomposes tasks into steps but then, in a tree-like fashion, explores multiple reasoning paths at once using techniques such as \gls{bfs}.

\subparagraph{Reflection} 
Self-reflection is a vital aspect that allows autonomous agents to improve iteratively by refining past action decisions and correcting previous mistakes. It is a common design pattern that involves the system explicitly criticizing and evaluating its output and subsequently refining it in an automated setup, often leading to notable performance gains.\autocite{shinn2023reflexion, madaan2023selfrefine, gou2024critic}

\subparagraph{Memory}
Another important building block for solving complex tasks is the ability to memorize information. 
Apart from the information provided to the agent within the context window of the \gls{llm}, it is possible to embed previous interactions of the agent in a vector database and retrieve them using \gls{rag}.\autocite{wang2023augmenting} 
This enables systems to retain and use information across very extended periods.

\subparagraph{Tool use}
As alluded to above, one of the most important design patterns for \gls{llm}-powered agents is to let them use external tools such as specialized models,\autocite{shen2023hugginggpt} web \glspl{api},\autocite{patil2023gorilla} simulation engines, or databases (e.g., citation traversal\autocite{skarlinski2024languageagentsachievesuperhuman}) to make up for the information and capabilities that might be missing. A powerful tool can also be additional \gls{llm} calls, as in the reranking and contextual summarization step in PaperQA.\autocite{skarlinski2024languageagentsachievesuperhuman} Here, the tool summarizes different chunks of papers and then rates the relevance for answering the question (in contrast to naïve retrieval that is typically performed in \gls{rag} systems).
The so-called ReAct framework is a common way to implement this.\autocite{yao2023react} It prompts the \gls{llm} to reason about the user's query and then act by choosing a tool. It is important to mention that the reasoning is done through inference, i.e., the model reasons only by the completion.
The actions will lead to observations (e.g., responses from an \gls{api} call) that might lead to further think-act-observe cycles before a final answer is returned.

A simple case about building an agent with custom tools can be found in \OnlineMaterial{Agents}{https://matextract.pub/content/agents/agent.html}.

\subparagraph{Multi-agent collaboration}
The multi-agent approach has proven to be a robust variant for self-reflection and self-criticism.\autocite{qian2023communicative} This approach involves making more than one agent work together. 
By doing this, it is possible to define different roles. For example, defining one evaluator agent and one critic agent can greatly improve the overall results. Another possibility is to include an agent that provides feedback or to add self-feedback to all the agents. Feedback can help the agents better understand possible areas of improvement, leading to better system performance. 
In multi-agent setups, it is even possible to create \enquote{creator} agents\autocite{talebirad2023multiagent} that create new agents with a specific role and goals to accomplish.

\paragraph{Limitations}
Current agent workflows are still limited in several ways, most importantly by the error rate of the base model, which leads to a high risk of failures in longer workflows.\autocite{ridnik2024codegenerationalphacodiumprompt} 
 
\subparagraph{Error amplification} 
A fundamental problem of autoregressive models such as current \glspl{llm} is that errors accumulate since the outputs of the models are the inputs for the subsequent generation.
In the case of agents, this means that in a long reasoning path, if one tool has an error, the agent passes this error to the next tool or reasoning step. The following tools can amplify the initial errors leading to large errors in the final answer.\autocite{song2023restgpt,zhuang2023toolqa} 

\subparagraph{Limited context-length} Similar to other applications of \glspl{llm}, agents are limited by the finite context length of the model.\autocite{andreas2022language} This is particularly pronounced for agent systems as they typically \enquote{memorize} the planning, reflection, or tool use traces within their finite context window.\autocite{shi2023large} This problem might be resolved by the rapidly growing context windows of frontier models. 

\subparagraph{Endless loop} Another situation that may happen is that the agents can end up stuck in the same loop forever. This situation can arise as a result of hallucinations and chaining different thoughts.\autocite{huang2024understanding} This is especially likely to happen when dealing with complex problems that the agent may not be able to solve.\autocite{talebirad2023multiagent}

\subparagraph{Safety-critical interactions} 
Since agents are supposed to interact with their environment,\autocite{cai2024largelanguagemodelstool, qian2024creatortoolcreationdisentangling, yuan2024craftcustomizingllmscreating} giving them access to computer systems or even laboratories requires additional safety precautions.\autocite{Boiko2023, ruan2024identifyingriskslmagents, tang2024prioritizingsafeguardingautonomyrisks}

\subparagraph{Challenges in evaluation} Finally, these agents are very difficult to evaluate and optimize since not only the final answer must be evaluated but also the reasoning process.\autocite{kapoor2024aiagentsmatter, huang2022language} 
The evaluation of the reasoning path can be necessary for the proper application of these systems for open-ended tasks as it allows for identifying the different sources of errors, i.e., if the error comes from a bad reasoning step of the \gls{llm} or if it comes from a wrong answer by one of the tools.\autocite{wang2023voyager}  Evaluating the reasoning path is a very difficult task because of the freedom that the agent has to reason and make decisions. In addition, most existing benchmarks rely on a ground truth; however, for many open-ended tasks, such a ground truth may not be available. 

\subsection{Postprocessing}
\subsubsection{Constrained decoding and enforcing valid outputs} \label{constrained_decoding}

\begin{figure}[htb]
    \centering
    \includegraphics[]{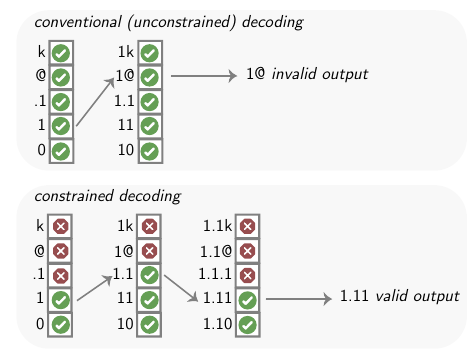}
    \caption{\textbf{Example of generating a number with conventional and constrained decoding.} In conventional decoding (top), all tokens have a non-zero probability of being sampled (indicated by green checkmarks). Thus, there is a non-zero probability that outputs are being generated that are not a number. With constrained decoding (bottom), however, we can dynamically adjust the set of tokens from which the system samples to generate only valid outputs. For instance, in the first step, only the two integer tokens are allowed. In the next step, also \texttt{.1} is allowed, as it might lead to a valid number. However, once \texttt{.1} has been sampled, it is no longer allowed (indicated by red crosses) as a number can only contain one decimal point.}
    \label{fig:constrained_decoding}
\end{figure}
The raw outputs of an \gls{llm} are probability distributions over tokens (see \Cref{sec:background}). 
The distributions indicate how likely each of these tokens (i.e., text fragments) are a continuation of the given text. By default, \glspl{llm} consider all possible tokens they have seen during training, which means every token has a nonzero probability.
There are different techniques (\enquote{sampling strategies}) for choosing the next token, thus text, from the probability distribution.
Naïvely, one might always choose the next token with the highest probability (\enquote{greedy decoding}). 
It has, however, been shown that this sometimes leads to unnatural text. Thus, practitioners often sample from the distribution to obtain natural text, where a factor called temperature indicates how frequently the sampling will give less probable tokens. 
This is nonideal for two reasons. First, we often know a priori that only a certain subset of tokens might be relevant---for example, we may care only about numbers, and limiting the pool of tokens we consider to only numbers will increase our chances of generating one. 
Second, sampling introduces randomness, which is typically unwanted in structured data extraction. The latter point can often be avoided (without taking into account randomness due to hardware or inherent tradeoffs in model deployment\autocite{puigcerver2023sparse}) by always taking the most probable token (i.e., setting temperature to zero). 
The former point can be implemented via so-called constrained decoding techniques. 
Those techniques can ensure that only a subset of \enquote{allowed} tokens will be used in the sampling stage (see \Cref{fig:constrained_decoding}). 

Constrained decoding can be implemented in different ways. One of the first widely popularized implementations has been in \texttt{jsonformer}.\autocite{jsonformer} 
The key to understanding this approach is that models generate outputs one token at a time. That is, predictions are used as inputs to make the next predictions (so-called autoregressive generation). 
If one aims to create structured data, one can make multiple optimizations in this process. First, some of these tokens are obvious, can be generated using code, and do not have to be generated by a model. 
For instance, code can generate much of the structure of a JSON file, such as the opening brace. 
That is, we can make the generation more efficient by only generating the content tokens using a model. 
Second, if we know the types of our data, we can limit the pool we sample from to only a subset of all tokens. For instance, \llamathree's vocabulary encompasses 128,256 tokens. If we want to fill a field with a boolean value, however, we only need to compare the values of the tokens for \enquote{True} and \enquote{False} (i.e., two out of 128,256 tokens).
In the simplest setting, one can start by already inserting a part of the desired output in the response, e.g., \texttt{\{"query":}. One can then decode until a stopping criterion is reached. This would be any token that is not a number or decimal separator (\texttt{.}) for simple numeric fields. From there, one can then start again by inserting all the text that is already known to be part of the desired output, e.g., a closing comma (\texttt{,}) and the next key.
These ideas can be extended to more complex constraining patterns, for example, based on formal grammars.\autocite{geng2024grammarconstrained}  
These formal languages can describe, in principle, the structure of any computable object.\autocite{Deutsch_2019} Thus, one could, for example, constrain the model to generate syntactically valid code in a programming language of choice (as programming languages can be described using formal grammar). 
While such constrained decoding is now well-supported in libraries such as \texttt{outlines},\autocite{willard2023efficient} \texttt{Instructor}\autocite{githubGitHubJxnlinstructor} (see \OnlineMaterial{Collecting data for reactions procedures}{https://matextract.pub/content/reaction_case/reaction.html}), \texttt{marvin},\autocite{githubmarbin} \texttt{ggml} (which supports grammars provided in Backus–Naur form),\autocite{githubGitHubJggml} or even the OpenAI \gls{api} (via JSON mode and function calling, or its newest feature, structured outputs, which ensures that the model's response always adheres to a provided schema), it is still not widely used for generative data extraction in the chemical sciences.\autocite{sayeed_2024} This is a promising future research direction as many relevant chemical datatypes (e.g., IUPAC names) can be represented with a formal grammar. One limitation with those approaches, however, is that they cannot naïvely be used with advanced prompting techniques that nudge the model to \enquote{think} by generating tokens. In those cases, it might make sense to use multiple prompts.

A middle ground between prompting without type and syntax constraints and constrained decoding can be to provide type hints in the prompt. These type hints can, for instance, also be literal, meaning a list of permitted strings. 

Examples of constrained decoding and enforcing valid outputs are shown in \OnlineMaterial{Constrained generation to guarantee syntactic correctness}{https://matextract.pub/content/constrained_decoding/index.html}.

To aid the integration with existing knowledge bases, one can also ground the LLM output on established ontologies as, for example, \textcite{Caufield_2024} have implemented in SPIRES, which demonstrated effectiveness in extracting chemical-disease relationships from biomedical literature, grounding entities to standardized identifiers (e.g., mapping \enquote{Cromakalim} to MESH:D019806).

\subsubsection{Evaluations}
\label{chap:evaluations}

\begin{figure}[htb]
    \centering
    \includegraphics[width=\textwidth]{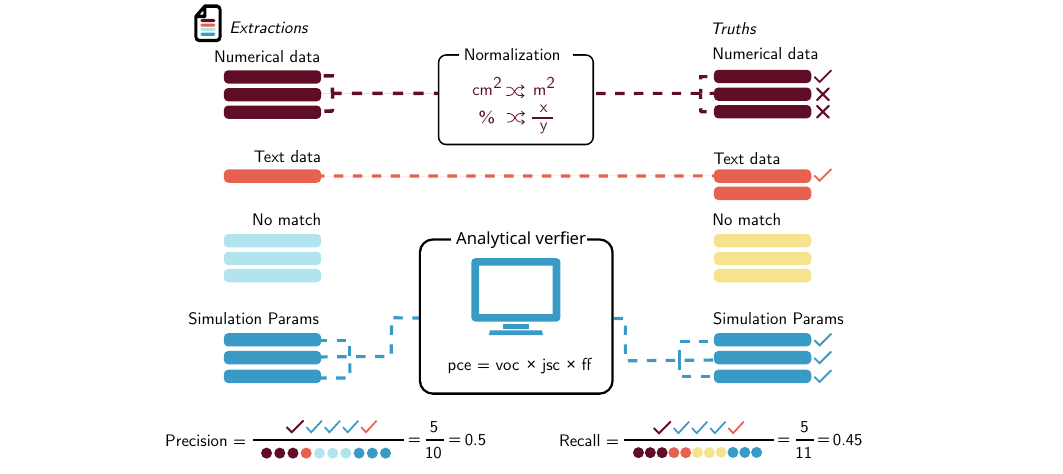}
    \caption{\textbf{Evaluations workflow.} The extracted structured data is displayed on the left, while the manually labeled truth data is on the right. The colors indicate matching on both sides. Checkmarks indicate correct keys, crosses indicate incorrect ones, and circles refer to the keys in the extracted and true data. Notice how there is an unmatched blue set in the extractions and a yellow set in the truths. This affects how precision and recall are calculated. Numerical data needs to be normalized if the units are reported differently from how they are stored in the truth data. Certain fields, like the simulation parameters, can be validated using scientific analysis tools to make sure they obey all domain rules.}
    \label{fig:evaluations}
\end{figure}

\noindent To optimize a system, its performance needs to be evaluated. In the case of structured data extraction, this is not trivial as there are many different yet related data items one extracts in a potentially nested data model. In addition, many entities can be reported in multiple synonymous ways. Thus, it is essential to carefully inspect the extracted data for common error sources. This manual inspection then allows an informed development of an automated evaluation pipeline that allows for systematic assessment of the extraction performance of different systems. As \textcite{sayeed_2024} indicate, the lack of a properly defined metric for structured data extraction makes systematic evaluation challenging.

\paragraph{Scoring the extraction of a simple entity extraction}

The simplest setting can be thought of as extracting a single value. For example, if the task is collecting all chemicals in a document, the performance of a model could be scored by measuring how many chemicals we found (retrieved entries) out of all the ones reported in the documents (relevant entries) and how many of the extracted chemicals are actually correct. The first measure is recall (\Cref{eq:recall}).

\begin{equation}
\text { recall }=\frac{\mid\{\text { relevant entries }\} \cap\{\text { retrieved entries }\} \mid}{\mid\{\text { relevant entries }\} \mid},
\label{eq:recall}
\end{equation}

where the $\cap$ symbol denotes a set intersection between the set of entries, and $\mid \mathrm{set} \mid$ indicates the number of items in a set.

The second measure is known as precision and is defined in almost the same way, except for the denominator being the retrieved entries (\Cref{eq:precision}).

\begin{equation}
\text { precision }=\frac{\mid\{\text { relevant entries }\} \cap\{\text { retrieved entries }\} \mid}{\mid\{\text { retrieved entries }\} \mid}.
\label{eq:precision}
\end{equation}

Intuitively, recall can be thought of as how many of the values we expected are extracted correctly and precision as out of just the extracted values how many are actually correct. Based on those two definitions, additional metrics can be defined, such as the $F_1$-score (which is the harmonic mean between precision and recall and also known as the Dice score). 

Due to the infancy of evaluations of structured extractions, there is much confusion about how to use metrics meant for classification models for this task. We propose to define \emph{True Positive} as a correct value extracted by the \gls{llm}. \emph{False Positive} is when there is an extraction but the value does not match what we expect. \emph{False Negative} can be taken as a value that is in our ground truth but has not been extracted by the \gls{llm}. \emph{True Negative} is not applicable and cannot be realistically defined for our task. True negative can be thought of as a value that we did not ask the \gls{llm} and it did not provide it. Therefore, it is every concept or word available in our document or vocabulary that hasn't been reported.
\Cref{tab:examples_error_types} provides examples of these different error types one might encounter in structured data extraction.
Using these definitions, we can calculate many metrics like the F$_1$-score as it does not use \gls{TN}. 
The recall and precision based on \gls{TP}, \gls{FP}, and \gls{FN} looks like this:

\begin{equation}
\text { precision }=\frac{\mid\text { \gls{TP} }\mid}{\mid\text { \gls{TP} } \mid \cap \mid\text { \gls{FP} } \mid}
\quad \text { recall }=\frac{\mid\text { \gls{TP} }\mid}{\mid\text { \gls{TP} } \mid \cap \mid\text { \gls{FN} } \mid}.\label{eq:recall2}
\end{equation}

\begin{table}
    \centering
    \caption{\textbf{Examples of error types.} In the case of true positives, the extracted data matches the ground truth labels. In case there has been an extraction, but for an entry that does not exist in the ground truth, we call it a false positive (\enquote{hallucination}). False negatives are values the \gls{llm} missed to extract. As there are potentially infinite true negatives (not existing data the \gls{llm} did not extract), it is not meaningful to consider them.} \label{tab:examples_error_types}
    \begin{threeparttable}
    \footnotesize

    \begingroup
    \setlength{\extrarowheight}{8pt}

\begin{tabularx}{\textwidth}{Xp{18em}XX}
    \toprule
    Outcome Type & Content Example & Extracted Data & Expected Data \\
    \midrule
    True Positive (TP) & The properties of TiO2 include a bandgap of 3.2 eV, which is typical for materials used in photocatalysis. & \textcolor[HTML]{e66150}{Formula: TiO2,} \textcolor[HTML]{5f0e25}{Bandgap: 3.2 eV} & \textcolor[HTML]{e66150}{Formula: TiO2,} \textcolor[HTML]{5f0e25}{Bandgap: 3.2 eV} \\
    \addlinespace[3mm]
    \specialrule{0.01pt}{0.1pt}{0.1pt}
    False Positive (FP) & The properties of TiO2 include a bandgap of 3.2 eV, which is typical for materials used in photocatalysis. & \textcolor[HTML]{e66150}{Formula: CeO2,} \textcolor[HTML]{5f0e25}{Bandgap: 3.37 eV} & \textcolor[HTML]{e66150}{Formula: TiO2,} \textcolor[HTML]{5f0e25}{Bandgap: 3.2 eV} \\
    False Positive (FP) & The properties of TiO2 include a bandgap of 3.2 eV, which is typical for materials used in photocatalysis. & \textcolor[HTML]{e66150}{Formula: TiO2,} \textcolor[HTML]{5f0e25}{Bandgap: 3.37 eV} & None \\
    \addlinespace[3mm]
\specialrule{0.01pt}{0.1pt}{0.1pt}
    False Negative (FN) & The properties of TiO2 include a bandgap of 3.2 eV, which is typical for materials used in photocatalysis. & \textcolor[HTML]{e66150}{Formula: TiO2,} & \textcolor[HTML]{e66150}{Formula: TiO2, Bandgap: 3.2 eV}  \\
    False Negative (FN) & The properties of TiO2 include a bandgap of 3.2 eV, which is typical for materials used in photocatalysis. & \textcolor[HTML]{e66150}{Formula: TiO2,} \textcolor[HTML]{81e0f5}{Mass: 2 g} & \textcolor[HTML]{e66150}{Formula: TiO2,} \textcolor[HTML]{f6ca15}{Bandgap: 1.02 eV} \\
    \addlinespace[3mm]
\specialrule{0.01pt}{0.1pt}{0.1pt}
    True Negative (TN)* & The properties of TiO2 include a bandgap of 3.2 eV, which is typical for materials used in photocatalysis. & None & None \\
    \addlinespace[3mm]
    \bottomrule
\end{tabularx}

    \endgroup

 \begin{tablenotes}
            \item[] *practically cannot be calculated.
        \end{tablenotes}
    \end{threeparttable}
    \label{tab:evaluation_scenarions}
    \normalsize
\end{table}

\paragraph{Matching to ground truth}
One challenge with these tailored metrics for more complex data models is the presence of multiple instances of the same object type. For instance, when extracting reactions, there may be more than one reaction in a paper. This poses a problem when scoring performance, as the ground truth and extracted output may contain different numbers of reactions. The model may have missed some reactions or falsely identified others. To accurately score a specific field, such as the yield of a reaction, the extracted entities must be matched with those in the ground truth before a comparison can be made.

To achieve this, it is recommended to first define a unique identifier for entities (such as a sorted list of the normalized reactant names in reaction) and then utilize a fuzzy matching score, such as the Levenshtein edit distance, to match each extracted entry with one element from the ground truth. 
This matching process can be seen as a one-to-one mapping between two lists (extracted outputs and ground truth), which is achieved by minimizing the total distance between all pairs. More formally, this is known as the linear sum assignment problem.\autocite{Burkard1980}

\paragraph{Data normalization}
For many extraction tasks, normalization is relevant before calculating metrics. For chemicals, this is relevant because there are often multiple equivalent ways of naming the same compound. While the normalization workflow will differ from use-case to use-case, tools such as PubChem\autocite{Kim_2015} or the Chemical Identifier Resolver Server\autocite{Sitzmann_2008} will often be useful to derive canonical representations for chemical compounds. Additionally, when dealing with units, tools such as \texttt{pint}\autocite{grecco_pint_2014} or \texttt{unyt}\autocite{Goldbaum2018} can be used to convert data into standard units before performing metric computations (see \OnlineMaterial{Evaluations}{https://matextract.pub/content/evaluations/evaluations.html}).

\paragraph{Overall metrics}
In addition to specific metrics, having a general metric, that provides a score on the whole extraction task,  can be helpful. The Damerau-Levenshtein edit distance is one such metric. It measures the minimum number of operations needed to change one document into another, including inserting, deleting, substituting, or transposing characters. This metric must be computed on canonicalized documents, i.e., documents that have been sorted and encoded similarly.

\paragraph{Validation using chemical knowledge and understanding} Data extraction in chemical and materials science has a significant advantage over other domains due to our understanding of rules and links between different data entries. This knowledge enables us to conduct \enquote{sanity checks} to ensure the accuracy and consistency of the extracted data. Despite this unique opportunity, it has not been widely utilized. A notable example is the work of \textcite{patiny_automatic_2023}, where they extracted molecular properties from text and used cheminformatics tools to validate the consistency. For instance, they extracted \gls{nmr} spectra and used cheminformatics tools to verify if they were consistent with the given molecular formula (see \OnlineMaterial{Validation case study: Matching NMR spectra to composition of the molecule}{https://matextract.pub/content/NMR_composition_matching/NMR_comp_matching.html} and  \OnlineMaterial{Retrieving data from chacolgenide perovskites}{https://matextract.pub/content/perovskite/constrained_formulas.html} for an example using stability criteria for perovskites). Similar validation can be performed on other experimental data, such as mass spectra and elemental analysis (see \OnlineMaterial{Collecting data for reactions procedures}{https://matextract.pub/content/reaction_case/reaction.html} for an example of checking the extraction of the correct number of atoms on both sides of a reaction). The benefit of these consistency checks is not only improving data quality but also enabling a first evaluation loop without the need for manually labeled data.

Beyond the use of validation based on chemical knowledge, it can also be practical to use another \gls{llm} to check, for example, for factual inconsistencies (e.g., if there were hallucinations during the extraction).

\section{Frontiers}
\label{sec:frontiers}
\glspl{llm} have greatly advanced the capabilities of data extraction given their ability to process text and other data modalities, such as figures through \glspl{vlm}. These advancements open up further compelling opportunities to make data extraction more robust and accessible, which we describe along a handful of research frontiers.  

\paragraph{Improving multimodal models} As described in \Cref{chap:multimodal_models}, data modalities beyond text often lead to unique challenges. While \glspl{vlm} can be quite helpful in extracting useful information from diverse types of data one might encounter in scientific literature, such as tables, formulas, structure files (e.g., CIF files for crystals \autocite{alampara2024mattext}), and sub-figures containing images with intricate relationships. Future work remains to make them more robust and amendable to the diversity of data found in materials science and chemistry. As described in \textcite{hira2024reconstructing}, some of the complexity can arise when diverse modalities are contained within a different data structure, such as chemical composition and related properties within a table that are linked to a figure in other parts of the document. Furthermore, the data format that current methods provide, such as \LaTeX\xspace or XML code, may not be ideal for training \glspl{llm} and other AI models for desired downstream tasks. As such, further work is needed to expand the capabilities of modern tools to not only provide structured data but also diverse forms of structured data.

\paragraph{Cross-document linking} As reported by \textcite{miret2024llms}, current extraction methods, including \glspl{llm} and \glspl{vlm},  mainly focus on extracting data contained within a single document or a sequence of documents given in a broader window, as described in \Cref{chunking}. Much of scientific data, however, relies heavily on referenced work for relevant concepts, descriptions, and experimental results (that might be in other repositories)\autocite{hira2024reconstructing, Ongari_2022}. In scientific publishing, including this work, it is very common for a scientific report to reference a procedure from another paper or report for brevity and accreditation. As such, the references and their potentially important context are usually not considered for information extraction methods today. The exclusion of such information likely limits the performance of modern \glspl{llm} and \glspl{vlm} to perform relevant chemical tasks \autocite{miret2024llms}. This also extends to data extraction, where understanding the relevant scientific background of a given figure, table, or data modality could help the \glspl{llm} process the data to the appropriate structured format and provide semantic meaning. One potential approach for addressing this challenge could be using multiple agents to analyze the same set of data with one agent providing relevant scientific background, such as a \gls{rag}-based chemistry \gls{llm}, and another providing extraction capabilities for the figure itself, such as \glspl{vlm}.

\paragraph{Scientific literature bias} The scientific literature is strongly biased towards positive results and highly refined information presented in curated text, figures, and tables. While it is important to maintain the high-quality standard in scientific publishing, there are likely adverse effects of not having \glspl{llm} observe negative results that are common in a wide range of real-world use cases, such as internal reports and communications \autocite{jia2019anthropogenic, raccuglia2016machine}. On top of that, many works of the scientific literature contain only incomplete information, which has fueled a reproducibility crisis in various fields, leading to concerns that advanced AI methods, such as \glspl{llm}, will make the situation worse \autocite{ball2023ai}. As such, it is important to improve data reporting methods that enable better dissemination of scientific knowledge in line with the development of scientific \glspl{llm}.

\paragraph{Beyond data extraction from papers}
Much of scientific innovation occurs when deploying new types of capabilities, such as synthesis equipment, characterization tools, and scientific simulation codes. The dynamic nature of these tools makes it possible for such new types of data structures and modalities to be continuously invented and deployed for diverse sets of applications. Ideally, \glspl{llm} should be capable of extracting relevant information from new tools and procedures. This represents a fundamentally new problem that future research work can tackle, building on the approaches described in this review. Agentic approaches described in \Cref{chap:agents} may be a useful framework given the flexibility of adding diverse tools for upcoming data modalities. 

\paragraph{From query to model}
Given the emergence of agentic systems that can autonomously build \gls{ml} models,\autocite{huang2024mlagentbenchevaluatinglanguageagents} it is not difficult to envision coupling data search agents (e.g., PaperQA\autocite{skarlinski2024languageagentsachievesuperhuman}) with data extraction agents and \gls{ml} agents. 
The result would be a system that can take in a search query and autonomously find data to train a model to answer the question. 
However, those systems would face the same challenges we discussed in \Cref{chap:agents} such as fragility and complex evaluation.

\paragraph{Benchmarks and open questions}
For information extraction, most existing benchmarks focused on evaluating the performance on separate tasks such as \gls{ner}, or \gls{re}.
More comprehensive benchmarks have been scarce, given the high cost of data labeling. This challenge, however, is unlikely to abate given the fact that new scientific discoveries continue to expand human scientific knowledge. As such, it might be useful to redirect the development of benchmarks towards adaptation of \glspl{llm} in low-data scenarios. While this has been observed in prior work, some potentially relevant benchmarks are too easy for modern \glspl{llm} \autocite{song-etal-2023-matsci} while others measure capabilities in adjacent tasks \autocite{wang2024scibench}. New benchmarks should address the challenges of \glspl{llm} and \glspl{vlm} for current capabilities and enable research along the current frontiers described above. This prompts a diverse set of remaining questions that the research community can work towards, including but not limited to:
\begin{itemize}
    \item What modalities exist in chemistry that current \glspl{llm} and \glspl{vlm} cannot process?
    \item What methods can appropriately handle the complex relationships of knowledge contained across multiple documents and knowledge sources in the chemical sciences?
    \item How can we mitigate scientific literature bias\autocite{jia2019anthropogenic} to build more comprehensive databases that can be digested by \glspl{llm}?
    \item How can we integrate the developments in \glspl{llm} for data extraction into broader efforts to build performant scientific assistants for chemistry?
\end{itemize}

\section{Conclusions}
 Structured data is immensely important for the advancement of science. 
 As shown in \Cref{fig:struc_vs_paper}, the aggregate information across diverse sub-fields in chemistry and materials science continues to grow at a notable pace. While there have been prior attempts to systematically extract data from these sources, only \glspl{llm} present a scalable solution to address both the breadth and scale of scientific data.

We hope that this review enables chemists and materials scientists to profit from these developments, 
 thereby accelerating the understanding and discovery of new compounds that further scientific knowledge and enable extraordinary technological advancement---leading from text to insights.

\section{Acknowledgement}

The research of M.S.-W.\ and K.M.J.\ was supported by the Carl-Zeiss Foundation as well as Intel and Merck via the AWASES programme. S.S., C.T.K., J.A.M., and K.M.J.\ are part of the NFDI consortium FAIRmat funded by the Deutsche Forschungsgemeinschaft (DFG, German Research Foundation) – project 460197019. M.R.G.\ and M.V.G.\ acknowledge financial support from the Spanish Agencia Estatal de Investigaci\'{o}n (AEI) through grants TED2021-131693B-I00 and CNS2022-135474, funded by MICIU/AEI/10.13039/501100011033 and by the European Union NextGenerationEU/PRTR, and support from the Spanish National Research Council (CSIC) through Programme for internationalization i-LINK 2023 (Project ILINK23047).
We thank Adrian Mirza and Maximilian Greiner for their feedback on a draft of this article.

\section*{Data availability}
The online book is available at \url{https://matextract.pub} for which the source code is available at \url{https://github.com/lamalab-org/matextract-book} and archived at \href{https://zenodo.org/records/14249541}{DOI 10.5281/zenodo.14249541}. 

\printnoidxglossary[type=\acronymtype]
\glsaddall
\printnoidxglossary[sort=letter]

\printbibliography

\end{document}